\documentclass[conference]{IEEEtran}
\IEEEoverridecommandlockouts

\usepackage{cite}
\usepackage{amsmath,amssymb,amsfonts}
\usepackage{algorithmic}
\usepackage{graphicx}
\usepackage{textcomp}
\usepackage{xcolor}
\def\BibTeX{{\rm B\kern-.05em{\sc i\kern-.025em b}\kern-.08em
    T\kern-.1667em\lower.7ex\hbox{E}\kern-.125emX}}

\usepackage{booktabs}        
\usepackage{tabularx}
\usepackage{threeparttable}  
\usepackage{siunitx}        
\usepackage[caption=false,font=footnotesize]{subfig}
\usepackage{tikz}
\usepackage{pgfplots}
\pgfplotsset{compat=1.18}
\usetikzlibrary{patterns}
\usepackage[table]{xcolor}
\usepackage{colortbl}
\usepackage{array}
\usepackage{enumitem}
\usepackage{makecell}
\newcommand{\StageCell}[2]{%
  \parbox[c][#1\baselineskip][c]{2.8cm}{\centering\bfseries #2}%
}

\newcommand{\ShadeCap}{60}
\newcommand{\scaleShade}[1]{\the\numexpr #1*\ShadeCap/100\relax}

\usepackage{url}

\begin{document}

\title{SoK: Stablecoins in Retail Payments
\thanks{*Yuexin Xiang is the corresponding author.}
}

\author{
Yuquan Li$^{1}$,
Yuexin Xiang$^{2,*}$,
Qin Wang$^{3}$,
Tsz Hon Yuen$^{2}$,
Andreas Deppeler$^{4}$,
Jiangshan Yu$^{5}$
\\[1ex]
\IEEEauthorblockA{$^{1}$ Faculty of Business and Economics, The University of Melbourne, Melbourne, Australia}
\IEEEauthorblockA{$^{2}$ Faculty of Information Technology, Monash University, Melbourne, Australia}
\IEEEauthorblockA{$^{3}$ CSIRO Data61, Sydney, Australia}
\IEEEauthorblockA{$^{4}$ School of Business, Monash University, Bandar
Sunway, Malaysia}
\IEEEauthorblockA{$^{5}$ School of Computer Science, The University of Sydney, Sydney, Australia}
}


\maketitle

\begin{abstract}
Stablecoins have emerged as a rapidly growing digital payment instrument, raising the question of whether blockchain-based settlement can function as a substitute for incumbent card networks in retail payments. This Systematization of Knowledge (SoK) provides a systematic comparison between stablecoin payment arrangements and card networks by situating both within a unified analytical framework. We first map their respective payment infrastructures, participant roles, and transaction lifecycles, highlighting fundamental differences in how authorization, settlement, and recourse are organized. Building on this mapping, we introduce the CLEAR framework, which evaluates retail payment systems across five dimensions: cost, legality, experience, architecture, and reach.

Our analysis shows that stablecoins deliver efficient, continuous, and programmable settlement, often compressing rail-level merchant fees and enabling 24/7 value transfer. However, these advantages are accompanied by an inversion of the traditional pricing and risk-allocation structure. Card networks internalize consumer-side frictions through subsidies, standardized liability rules, and post-transaction recourse, thereby supporting mass-market adoption. Stablecoin arrangements, by contrast, externalize transaction fees, error prevention, and dispute resolution to users, intermediaries, and courts, resulting in weaker consumer protection, higher cognitive burden at the point of interaction, and fragmented acceptance. Accordingly, stablecoins exhibit a conditional comparative advantage in closed-loop environments, cross-border corridors, and high-friction payment contexts, but remain structurally disadvantaged as open-loop retail payment instruments.

\end{abstract}


\section{Introduction}

Retail payments are end-user initiated transfers for routine commercial and personal purposes and are widely regarded as essential infrastructure for contemporary economic activity~\cite{BIS2003,hasan2012retail}. They encompass the arrangements, instruments, and technical infrastructures that enable households and firms to initiate, authorize, and settle low-value transactions at scale. By reliably linking consumers, merchants, and financial intermediaries, these systems reduce checkout frictions and transaction costs while sustaining confidence in everyday exchange~\cite{ECB2010_PaymentSystem,BIS2003}. Continued improvements in retail payment systems have further supported economic performance. Empirical evidence suggests a strong correlation between the adoption of efficient electronic instruments and higher economic output, consumption, and trade, particularly in markets where harmonization enhances system performance~\cite{hasan2013retail}.

Retail payments in mature payment markets are dominated by card network payments, including card-present transactions and transactions initiated via mobile wallets~\cite{lagator2021global}. Traditional card schemes, such as Visa and Mastercard, offer sub-second authorization, high reliability, and predictable settlement through standardized messaging and interbank clearing in central bank money~\cite{BIS2016, ECB2020}. These systems have matured into a highly coordinated ecosystem of issuers, acquirers, and processors, bolstered by regulatory frameworks such as the Revised Payment Services Directive (PSD2)~\cite{yawe2020impact}, which ensures competition, security, and consumer recourse.

\begin{figure*}[htbp]
  \centering
  \includegraphics[width=0.85\linewidth]{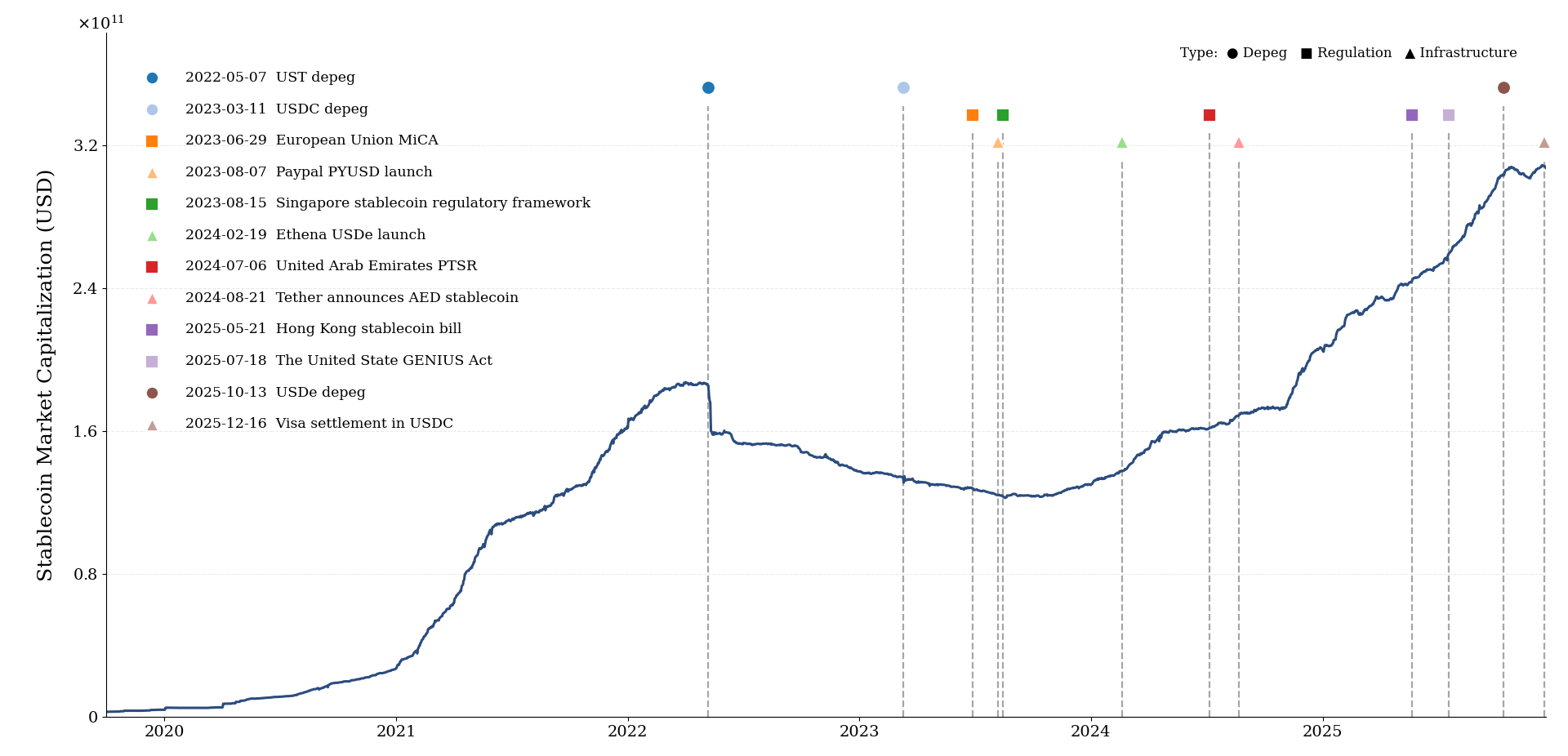}
  \caption{Global stablecoin market capitalization with major depeg, regulatory, and infrastructure events highlighted}
  \label{fig:stablecoin-mcap}
\end{figure*}

However, a challenger ecosystem has emerged. Instead of the high-volatility assets that defined the early cryptocurrency market, stablecoin, a digital asset intended to maintain a stable value relative to a reference fiat currency through reserve or algorithmic models, has scaled rapidly as a potential alternative payment rail. By October 2025, the stablecoin supply exceeded \$300 billion~\cite{defillama2025} as shown in Fig.~\ref{fig:stablecoin-mcap}, with on-chain settlement volumes reaching approximately \$8.9 trillion in the first half of the year~\cite{Rise2025Stats}. At the application layer, the convergence of Web3~\cite{wang2022exploring} and retail is accelerating. For example, the integration of OKX Pay’s stablecoin with GrabPay in Singapore demonstrates that stablecoin rails are increasingly being integrated into familiar Near Field Communication (NFC) and QR-based checkout flows~\cite{okx2025pay,businesstimes2025okx,cna2025xsgd}.

Stablecoins differ from other crypto-assets in that they are explicitly designed to maintain a stable value against a reference fiat currency, making them suitable for transactional use in retail payments. Their widespread adoption as settlement assets across exchanges and Decentralized Finance (DeFi) platforms~\cite{tian2021enabling,brechlin2024buy,werner2022sok,zhou2023sok} demonstrates operational scalability, even though these contexts differ from point-of-sale commerce. Recent regulatory initiatives, including the Markets in Crypto-assets Regulation (MiCA)~\cite{micaReg} and the GENIUS Act~\cite{US2025GENIUS}, further clarify issuance and conduct requirements, strengthening their relevance as regulated payment instruments. These features make stablecoins the most appropriate blockchain-based basis for comparison with card- and account-based payment networks.

Despite rapid growth in cross-border and institutional applications, the extent to which stablecoins can achieve functional parity with card networks at the retail Point of Sale (POS) remains uncertain. Prior studies highlight several dimensions along which decentralized payment rails differ structurally from centralized card networks. First, blockchain settlement does not always confer the same legal finality as card-scheme payments, and transaction reversibility is not natively supported at the protocol layer~\cite{neyer2017blockchain,koutrouli2025exploring}. Second, operational complexity continues to constrain merchant adoption, as access to crypto-fiat conversion, refund handling, and reconciliation remains less standardized than the workflows provided by traditional acquirers~\cite{erdin2021evaluation, jonker2019drives}. Third, responsibility within stablecoin arrangements is distributed across different entities, including issuers, wallet providers, and infrastructure operators, while the underlying blockchains themselves do not bear contractual liability~\cite{nabilou2020dark, puschmann2024taxonomy}. Finally, systematic comparison is challenged by the absence of harmonized performance statistics, requiring analytical frameworks that can organize heterogeneous evidence and metrics across payment architectures~\cite{worldbank2018gpss,ECB2020,ArnerAuerFrost2020}.

While prior research has extensively analyzed stablecoins, existing Systematization of Knowledge (SoK) studies primarily examine them through design, economic, or security lenses, rather than as retail payment infrastructures. Early taxonomies such as Moin et al.~\cite{moin2019sok} classify stablecoins by stabilization and collateral mechanisms, while broader surveys~\cite{mahrous2025sok} integrate economic, regulatory, and technical perspectives without evaluating transaction-level payment performance. More recent security-focused frameworks~\cite{ling2025sok} further concentrate on smart-contract and composability risks in DeFi. Consequently, the suitability and performance of stablecoins as retail payment rails remain insufficiently examined in prior SoK studies.

To address these gaps, this study compares stablecoin arrangements with card networks through a unified analytical perspective. It anchors this comparison in the \textbf{CLEAR} framework, which systematizes the evaluation of \underline{C}ost, \underline{L}egality, \underline{E}xperience, \underline{A}rchitecture, and \underline{R}each. 

The analysis focuses specifically on \emph{asset-backed} stablecoins, which dominate current retail-oriented designs and provide the most direct counterpart to incumbent payment systems; algorithmic stablecoins are excluded because their historical instability makes them unsuitable as a basis for retail payment infrastructure (as the UST collapse noted in Fig.~\ref{fig:stablecoin-mcap}). Based on this framework, we address the following research questions (RQs):
\begin{itemize}
\item \textbf{RQ1:} What are the core instruments, technical rails, and participant roles that differentiate stablecoin payment systems from card networks?
\item \textbf{RQ2:} How do stablecoin payment systems perform relative to card systems when assessed through the CLEAR framework?
\item \textbf{RQ3:} 
What institutional, governance, and interoperability conditions are required for stablecoins to match the levels of transaction assurance provided by card networks and enable large-scale retail adoption?
\end{itemize}

\smallskip
\noindent\textbf{Contributions.} Having established the analytical focus, this study advances three primary contributions:
\begin{itemize}
\item We establish a common basis for comparison by aligning the transaction lifecycles of stablecoin-based payments and card networks.
\item We develop and apply the CLEAR framework to synthesize evidence on the strengths and limitations of both payment architectures.
\item We propose a measurement framework for assessing the conditions that would allow stablecoins to attain parity with card systems in retail payment.
\end{itemize}

Throughout this paper, the term \emph{card networks/systems} refers to the underlying open-loop payment rails such as Visa that govern authorization, clearing, and settlement. Mobile wallets such as Apple Pay, as well as open-loop transportation payments that route transactions over these networks, are treated as user-facing interfaces within this architecture rather than as independent payment infrastructures.

The remainder of this paper is organized as follows. Section~\ref{sec:mapping} outlines the retail payment landscape and establishes the basis for comparing stablecoin and card network models. Section~\ref{sec:clear} applies the CLEAR framework to evaluate their performance across five dimensions. Section~\ref{sec:discussion} synthesizes these findings and sets out implications for stablecoin progress toward retail parity. Section~\ref{sec:conclusion} summarizes our work.

\section{Mapping Payment System Landscape}
\label{sec:mapping}

Comparison between stablecoin-based payments and card networks requires clarification of the payment infrastructures on which they operate. Differences in settlement architecture, governance, and transaction lifecycles shape core performance outcomes. Accordingly, we begin by mapping the evolution of retail payment infrastructures and situating both systems within a common analytical framework.

\subsection{Payment Infrastructure}  
Retail payment systems have evolved from physical token exchange to account-based arrangements and, more recently, toward programmable value transfer. Economic research characterizes this evolution as a shift from token-based to account-based systems, reflecting persistent trade-offs among finality, scalability, and geographic reach~\cite{kahn2009pay}. Next, we present their evolution over time.

\smallskip
\noindent\textbf{Cash and Paper Payments.}
Early retail payment systems relied on physical tokens such as specie and cash and paper-based instruments such as cheques and giros. Cash offered immediate settlement finality in face-to-face settings but was constrained by physical materiality. Paper instruments allowed for remote payment but were operationally intensive: manual clearing processes produced settlement lags of several days and higher error rates~\cite{BIS2003}. Consequently, payment processing in this era was slow, locally fragmented, and costly for both banks and merchants.

\smallskip
\noindent\textbf{Electronic Card Payments.}
Since the 1970s, retail payments have increasingly migrated toward electronic, account-based infrastructures through card networks. Magnetic-stripe and later EMV cards established the four-party model with real-time authorization at the POS and deferred interbank settlement under standardized messaging (e.g., ISO~8583)~\cite{ward2006emv, stearns2011electronic,ISO_8583_2023,evans2004paying,BIS2016}. Subsequent digital and mobile innovations in the 2010s, including tokenization and biometric authentication in wallets such as Apple Pay, improved security and user experience but largely preserved existing rails: authorization is near-instant, while merchant settlement remains batch-based, typically occurring one or two business days later~\cite{ECB2020,thomadakis2022shortening,benito2025navigating}.

\smallskip
\noindent\textbf{Real-Time and Blockchain Payments.}
In response to the latency of legacy cards and ACH rails, the payment landscape has bifurcated into two modernizing paths: central bank-coordinated real-time systems and decentralized blockchain infrastructures.
\begin{itemize}
    
\item \textit{Real-Time Payment Systems.} Many jurisdictions have introduced Real-Time Payment (RTP) systems (e.g., SEPA Instant~\cite{epc_sctinst}, PIX~\cite{bcb_pix}, FedNow~\cite{frb_fednow}) that enable 24/7 interbank settlement in central-bank money, substantially reducing settlement delays~\cite{BIS2021,WorldBank2022}. While RTP improves the speed of legacy bank-based payment systems, integration into merchant checkout flows and card-like governance remains uneven.

\item \textit{Blockchain Settlement.} In parallel, blockchain technology introduced a distinct settlement architecture. Bitcoin demonstrated peer-to-peer value transfer without a central clearinghouse using cryptographic consensus to prevent double spending~\cite{nakamoto2008bitcoin}. Ethereum later extended this model by embedding smart contracts, enabling programmable and conditional transfers at the settlement layer~\cite{Buterin2014,zheng2020overview,garg2022ethereum}.

\item \textit{Stablecoin Payments.} Cryptocurrency volatility limited early blockchain assets as retail payment instruments~\cite{mattke2020cryptocurrency}, motivating the development of stablecoins. Stablecoins combine blockchain-based programmability with a price peg to fiat currencies, supported by heterogeneous backing assets~\cite{mahrous2025sok,TetherTransparency2025,CircleTransparency2025}, aiming to deliver monetary stability alongside global, continuous settlement~\cite{ArnerAuerFrost2020}.
\end{itemize}

This convergence makes the stablecoin model a natural point of comparison with the card network model. The two models are presented in sequence below.

\begin{figure}[htbp]
  \centering
  \includegraphics[width=0.8\linewidth]{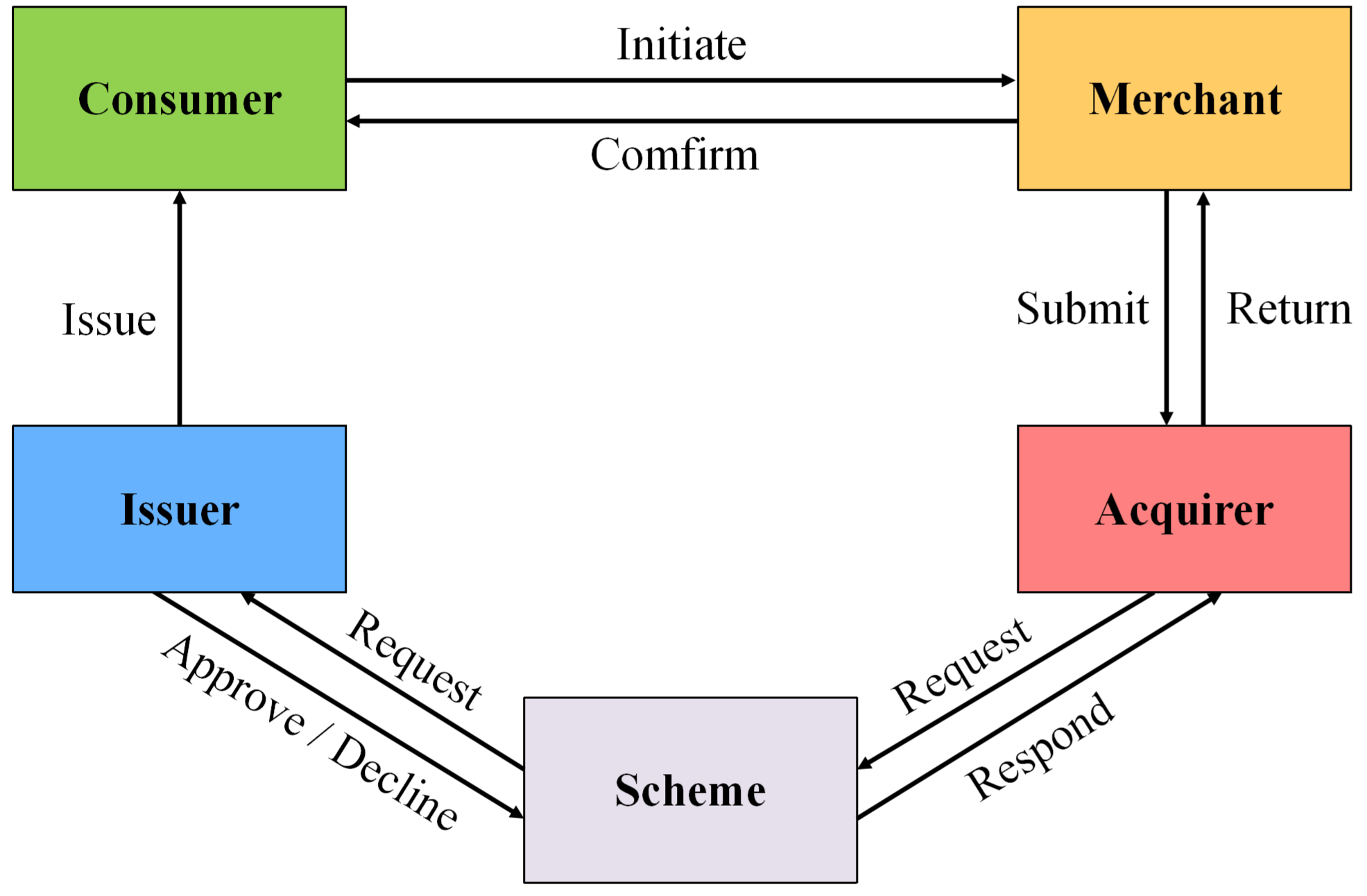}
  \caption{Card network model} 
  \label{fig:traditional-participants}
\end{figure}

\subsection{Card Network Model (Fig.~\ref{fig:traditional-participants})}\label{subsec:cardmodel}
Card payments can be characterized as a rule-based and multi-sided payment arrangement linking the cardholder, merchant, issuer, and acquirer, coordinated through a card scheme that standardizes messaging, risk management, and settlement procedures~\cite{alexandrova2009impact}. The arrangement operates within an account-based ledger structure in which transaction authorization verifies credentials and funds availability, while the transfer of monetary value is completed through subsequent clearing and interbank settlement rather than at the moment of authorization~\cite{cpmi2016glossary}.

\smallskip
\noindent\textbf{Instruments.} 
In retail payments, mainstream instruments can be grouped into four categories that are commonly used in official payment statistics and monitoring frameworks of central banks and standard-setting bodies: card payments, credit transfers, direct debits, and prepaid balances~\cite{hughes2010developments,humphrey201910}.

\begin{itemize}
  \item \textit{Card Payments}~\cite{rysman2007empirical}. Payments initiated using a physical or virtual card and routed through an open-loop card scheme. The underlying funds may be held in a bank deposit account or in an e-money account; however, the transaction is classified as a card payment because the card scheme rail is used at the POS. Under the four-party model, the issuer returns an authorization in seconds, while clearing and settlement occur subsequently under scheme rules and chargeback frameworks~\cite{worldbank2018gpss}. Examples include Visa, Mastercard, and UnionPay~\cite{aysan2025local}.
  
  \item \textit{Credit Transfers}~\cite{ECBGlossary}. Payments in which the payer instructs a Payment Service Provider (PSP) to push funds directly to the payee’s account over batch or instant rails. These instruments are widely used for bill payments and payroll, and, where instant rails and request-to-pay overlays exist, they can also support POS and e-commerce transactions. Examples include SEPA credit transfer~\cite{epc_sctinst}, ACH, faster payments~\cite{payuk_fps}, and RTP.
  
  \item \textit{Direct Debits}~\cite{MoneySmartDirectDebits}. Payments initiated by the payee, based on a prior mandate, to pull funds from the payer’s account on agreed dates. Scheme rulebooks govern mandate management, notifications, and dispute and refund rights, making direct debits the standard instrument for recurring payments such as utilities and subscriptions. Examples include SEPA direct debit~\cite{epc_sdd} and Bacs direct debit~\cite{bacs_dd}.
  
  \item \textit{Prepaid Payments}~\cite{lilge2001evolution}. Payments executed using prepaid value stored with an issuer and transferred within the issuer’s proprietary closed-loop infrastructure, rather than over open-loop card schemes or interbank transfer rails. The stored value represents a claim on the issuer and is accessed within the system’s own payment environment. Examples include transit cards and retailer gift cards~\cite{horne2016gift}.

\end{itemize}

\smallskip
\noindent\textbf{Participants.}
Card networks are commonly described using a four-party model~\cite{prager2009interchange}, consisting of the consumer, the merchant, the issuer, and the acquirer. In practice, these bilateral relationships are coordinated through a card scheme that provides the rulebook and network infrastructure linking issuers and acquirers. Fig.~\ref{fig:traditional-participants} summarizes these roles and the associated transaction and fee flows. The arrangement therefore involves the following participants~\cite{chakravorti2003theory, scardovi2017digital}:

\begin{itemize}
    \item \textit{Consumer.} 
    The transaction is initiated by the consumer, who presents a payment card at the merchant’s POS or checkout interface. This action signals consent to transact and triggers the creation of a payment request addressed to the merchant. Upon receiving the authorization outcome relayed back through the network, the consumer receives confirmation of approval or decline. Economically, authorization commits the issuer to honor the payment, with actual monetary settlement deferred beyond the point of interaction.
    
    \item \textit{Merchant.} 
    The merchant is the retailer selling goods or services~\cite{giardina1993merchant}. Upon receiving the consumer’s payment initiation, the merchant submits the transaction details to its acquirer. This submission represents a pull request for funds, whereby the merchant asks the acquirer to obtain authorization and, ultimately, settlement from the issuer. After authorization is completed and the response is returned, the merchant proceeds with the delivery of goods or services. Depending on scheme rules, subsequent disputes may lead to reversal after authorization.

    \item \textit{Acquirer.} The acquirer (e.g., Worldpay~\cite{worldpay}, Stripe~\cite{stripe_au}, and Adyen~\cite{adyen_au}) is the merchant-facing financial institution or processor that receives transaction submissions from merchants and forwards authorization requests into the card scheme network. Acting as the gateway to the scheme, it formats, validates, and routes transaction data, relays the authorization response back to the merchant, and subsequently participates in the settlement process under scheme rules. In exchange for these services, the acquirer charges the Merchant Discount Rate (MDR).~\cite{guo2012interchange}.

    \item \textit{Issuer.} 
    The issuer is the consumer’s bank and the ultimate decision-maker in the authorization process. After receiving the authorization request from the scheme, the issuer evaluates the transaction using account balance checks and fraud detection algorithms, then returns an approval or decline message formatted according to ISO 8583~\cite{ISO_8583_2023}. The issuer maintains the ledger of the consumer’s deposit balance or credit line and assumes both credit risk, when cardholders fail to repay, and fraud risk, when unauthorized transactions occur. These services are compensated through interchange fees received from the acquirer.

    \item \textit{Scheme.} 
   The scheme (e.g., Visa, Mastercard, and UnionPay~\cite{unionpay_intl}) operates the central routing and clearing infrastructure that connects acquirers and issuers. It forwards authorization requests from acquirers to issuers and relays issuer decisions back along the same path, while defining technical and operating standards and ensuring interoperability across participants. These functions are funded through scheme fees charged to participating institutions~\cite{rba_scheme_fees}.
\end{itemize}

\begin{figure*}[htbp]
  \centering
  \includegraphics[width=0.95\linewidth]{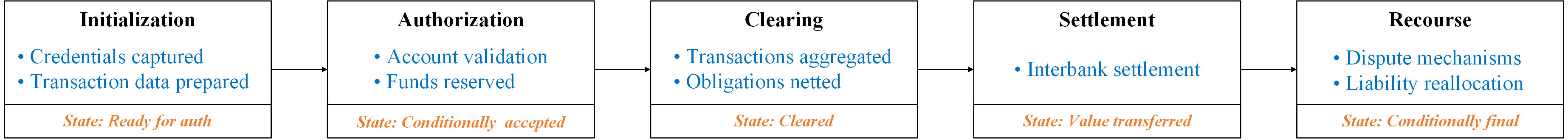}
  \caption{Lifecycle of card network model} 
  \label{fig:life-card}
\end{figure*}

\smallskip
\noindent\textbf{Transaction Lifecycle.} 
The card network lifecycle can be decomposed into five sequential phases as shown in Fig.~\ref{fig:life-card}, highlighting the temporal separation between information processing and liquidity transfer~\cite{worldbank2018gpss}.

\begin{enumerate}
  \item \textit{Initialization.} A transaction is initiated at the Point of Interaction (POI)~\cite{verifone_poi}, where payment credentials are captured and transaction data are prepared for network processing~\cite{stripe2023}. At this stage, the system establishes the parameters required for subsequent authorization, but no account validation or value transfer occurs.
  \item \textit{Authorization.} The authorization stage involves real-time account validation by the issuer. Upon receiving an authorization request routed through the acquirer and card network using standardized messaging formats such as ISO~8583~\cite{ISO_8583_2023}. The issuer verifies card validity, available funds or credit, account status, and fraud controls, and returns an approval or decline decision within seconds~\cite{visaCoreRules2025, checkout2024_cardAuthorization}. An approved authorization results in a temporary hold on the cardholder’s account, creating a conditional payment obligation under scheme rules~\cite{visaCoreRules2025}. At this stage, no interbank transfer of liquidity occurs; authorization is informational and contractual in nature, distinct from clearing and settlement, which occur subsequently~\cite{worldbank2018gpss,cpmi2016glossary}.
  \item \textit{Clearing.} Clearing is the post-authorization process in which approved transactions are aggregated and reconciled by the card network, and net obligations between issuers and acquirers are calculated under scheme rules to form settlement positions, without yet transferring funds~\cite{cpmi2016glossary,worldbank2018gpss}.
  \item \textit{Settlement.} Settlement is the stage at which payment obligations are discharged through the transfer of funds between parties~\cite{cpmi2016glossary}. In card payment systems, settlement typically occurs after clearing with a delay (often T{+}1 or T{+}2)~\cite{thrivent2025}, reflecting deferred net settlement arrangements and liquidity management practices rather than real-time value transfer~\cite{thrivent2025}. At this stage, net positions between issuers and acquirers are settled using interbank money, most often through central bank settlement infrastructures such as Real-Time Gross Settlement systems (RTGS), completing the monetary leg of the transaction~\cite{BIS2003}.
  \item \textit{Recourse.} Distinctively, the card payment lifecycle remains open after settlement. Card scheme rules provide post-settlement dispute and chargeback mechanisms that allow issuers to reclaim funds from acquirers and merchants in cases such as fraud or non-delivery~\cite{visaCoreRules2025}. While interbank settlement is final at the infrastructure level, these contractual mechanisms permit subsequent reallocation of funds, rendering settlement economically final but legally conditional for end users~\cite{kahn2002economics}.
\end{enumerate}

\begin{figure}[t]
  \centering
  \includegraphics[width=0.8\linewidth]{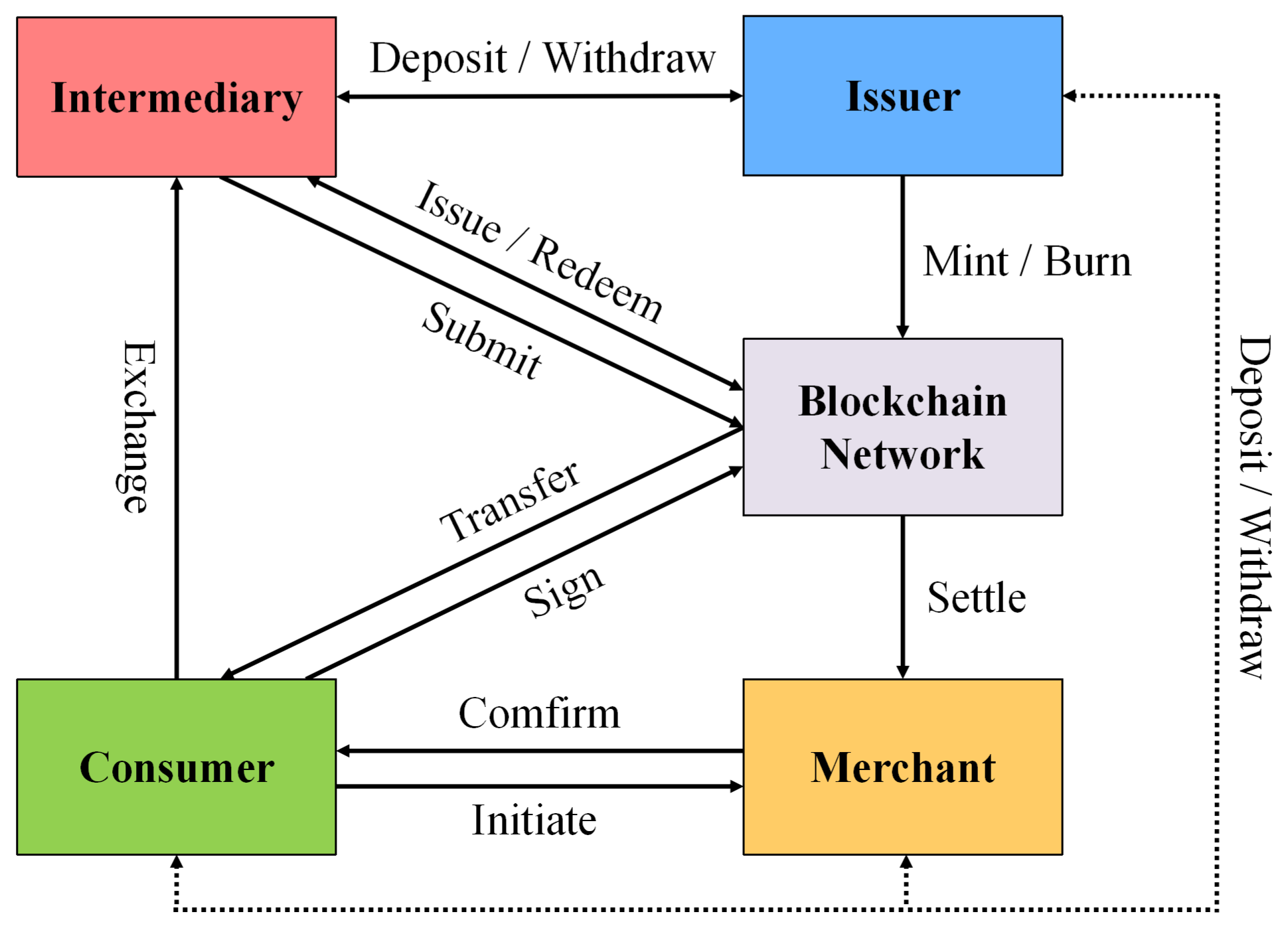}
  \caption{Stablecoin model} 
  \label{fig:stablecoin-participants}
\end{figure}

\subsection{Stablecoin Model (Fig.~\ref{fig:stablecoin-participants})}
\label{subsec:cryptomodel}

\smallskip
\noindent\textbf{Instruments.}
Stablecoins constitute the primary cryptocurrency payment asset for retail and on-chain settlement due to their price stability relative to fiat units of account~\cite{RBA2022}. However, stablecoin transactions do not operate in isolation: their issuance, transfer, and programmability are intrinsically dependent on other classes of crypto assets that support execution and protocol-level functionality~\cite{werner2022sok}.

\begin{itemize}
  \item \textit{Payment Layer: Stablecoins.}
  Stablecoins are digital tokens designed to maintain a stable value relative to a reference asset~\cite{US2025GENIUS}, typically a fiat currency. Their price stability makes them suitable as a medium of exchange and unit of account for retail payments and on-chain settlement.

  \item \textit{Execution Layer: Native Chains.} Stablecoin issuance, transfer, and settlement are typically executed on public blockchains~\cite{ferdous2021survey} (e.g., Ethereum and Solana~\cite{yakovenko2018solana}). These native chains provide consensus, transaction ordering, and finality~\cite{asayag2018fair,das2019transaction}, while their associated native tokens are used to pay transaction fees and incentivize validators. As a result, the cost, latency, and reliability of stablecoin payments are directly shaped by the performance characteristics and congestion dynamics of the underlying chain.

  \item \textit{Functionality Layer: Protocols.} Within DeFi environments, stablecoins interact with smart contracts deployed on application protocols such as decentralized exchanges (DEXs)~\cite{malamud2017decentralized}, lending platforms~\cite{kaplan2023decentralized}, and bridges~\cite{li2025blockchain}. These protocols are commonly associated with utility tokens that confer governance rights, access to protocol services, or eligibility for liquidity incentives. While such tokens are not used as means of payment at the POS, they support the circulation, composability, and programmability of stablecoins across applications.
\end{itemize}

\smallskip
\noindent\textbf{Participants.}
Stablecoin-based payment arrangements diverge from the structure of card networks. Rather than relying on a centralized issuer-acquirer-scheme hierarchy with contractual clearing and settlement obligations, stablecoin payments operate through a modular architecture composed of customers, issuers, blockchain networks, merchants, and intermediaries as shown in Fig.~\ref{fig:stablecoin-participants}. 

Each participant substitutes for a layer of the card payment stack, while reallocating authorization, settlement, and risk management functions from institutions to software and market-based intermediaries. This functional reallocation creates a responsibility vacuum: unlike the hierarchical card model in which the scheme acts as the ultimate contractual arbiter, the modular structure of stablecoin payments lacks a single entity that is accountable for end-to-end transaction success.

\begin{itemize}
    \item \textit{Customer.}
    The customer initiates stablecoin payments using a wallet~\cite{suratkar2020cryptocurrency}, which constructs and signs transfer instructions and submits them to the blockchain network for execution~\cite{dotan2021survey}. Payment is authorized by the user’s signature, and settlement occurs directly on the ledger as part of transaction execution. Custodial wallets operated by intermediaries perform compliance checks, transaction monitoring, and internal ledgering~\cite{kazerani2017determining,trautman2015commerce,chen2020you}, whereas self-custodial wallets provide direct network access while placing operational responsibility on the user~\cite{lee2023using,PhantomWallet2025}.

    \item \textit{Issuer.}
    Stablecoin issuers such as Circle~\cite{Circle2025} and Tether~\cite{Tether2025}, which issue USDC and USDT respectively~\cite{qin2021cefi}, are responsible for minting and burning tokens in response to fiat deposits and redemptions~\cite{arslanian2022stablecoins}. Supply changes are effected on-chain through mint and burn transactions, while fiat settlement and custody occur off-chain via intermediaries. Unlike card issuers, stablecoin issuers do not authorize retail transactions or extend credit at the POS; their role is limited to maintaining reserve backing and ensuring redemption at par.

    \item \textit{Blockchain Network.}
    The blockchain network serves as the shared execution and settlement infrastructure of stablecoin payments, maintaining the canonical ledger, validating transactions, enforcing consensus rules, and producing settlement finality~\cite{xu2023survey,nabilou2024probabilistic}. Signed transactions are submitted, ordered, executed, and finalized on-chain, resulting in balance updates whose latency and finality depend on the underlying consensus design~\cite{gramoli2020blockchain,xiao2020survey}. Unlike card schemes, the blockchain network provides no contractual guarantees, chargeback mechanisms, or dispute resolution; settlement is technical rather than legal.

    \item \textit{Merchant.}
    Merchants accept stablecoin payments by receiving on-chain transfers and monitoring network confirmations. Merchant-facing software providers may assist with transaction detection, pricing, and reconciliation, but they do not assume the risk-sharing role of acquirers. Once a transaction is finalized on the blockchain, it becomes effectively irreversible, and merchants do not benefit from scheme-level fraud protection or chargeback rights~\cite{politou2019blockchain,das2019transaction,ye2023survey}. Thus, fraud and operational risk are borne more directly by merchants and customers than in card systems.

    \item \textit{Intermediary.}
    Intermediaries, most prominently crypto exchanges~\cite{johnson2020decentralized,okasova2024using}, serve as the primary interface between the fiat monetary system and the stablecoin ecosystem. They aggregate issuance and redemption demand on behalf of users and merchants and provide fiat-stablecoin conversion, interacting with issuers off-chain and submitting mint and burn instructions to the blockchain network~\cite{CircleUSDC2025,TetherRedeem2025}. Functionally, intermediaries combine elements of acquirers and settlement banks in card networks, but operate without centralized scheme rules or mandated liability frameworks, making them critical for liquidity and the practical usability of stablecoins in retail payments.
\end{itemize}

\begin{figure*}[htbp]
  \centering
  \includegraphics[width=0.95\linewidth]{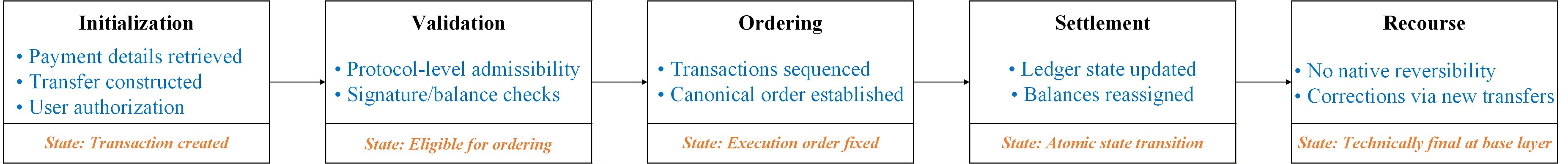}
  \caption{Lifecycle of stablecoin model} 
  \label{fig:life-coin}
\end{figure*}

\smallskip
\noindent\textbf{Transaction Lifecycle.} 
The stablecoin payment lifecycle can be organized into five phases as shown in Fig.~\ref{fig:life-coin}, reflecting how transaction creation, admissibility, ordering, and state transition are executed within a unified blockchain ledger. Unlike card systems, these functions are integrated at the protocol level, while the effective acceptance of payment depends on the underlying consensus design and confirmation policy.

\begin{enumerate}
    \item \textit{Initialization.} A retail stablecoin payment begins when the consumer’s wallet retrieves the merchant’s receiving address, and constructs a transfer instruction. The user authorizes this instruction through a cryptographic signature~\cite{houy2023security}. In custodial settings, authorization may occur as an internal approval, with the provider signing and submitting the transaction on the user’s behalf.

    \item \textit{Validation.} The transaction is propagated to the network, where nodes verify protocol-level admissibility, including signature correctness, nonce sequencing, and balance sufficiency~\cite{puthal2018everything,lashkari2021comprehensive}. In parallel, custodial intermediaries may apply additional risk or compliance screening prior to submission. This stage determines whether a transaction is eligible for execution, rather than whether payment is economically accepted.

    \item \textit{Ordering.} Admissible transactions are ordered into blocks or execution batches~\cite{ruan2020transactional,salau2022making}. In contrast to deferred, end-of-day clearing in card schemes, ordering in blockchain networks occurs continuously, establishing a canonical transaction sequence that directly determines ledger state transitions.

    \item \textit{Settlement.} Upon acceptance of a block under the consensus mechanism, the ledger state is updated and balances are reassigned~\cite{zhang2020overview,zhang2021transaction}. In this sense, clearing and settlement are collapsed into a single atomic state transition. However, the strength of finality depends on the consensus model and confirmation depth, particularly under probabilistic-finality designs. Merchants therefore typically treat payment as complete only after a predefined confirmation policy is satisfied.

    \item \textit{Recourse.} Because base-layer transfers rely on immutable ledgers and lack a network-wide adjudication authority, stablecoin payments do not support native chargebacks or scheme-mandated reversals~\cite{politou2019blockchain,asante2021distributed}. Corrective action generally requires a new transaction initiated by the recipient. While smart contracts or custodial platforms may implement escrow, dispute resolution, or rollback-like controls~\cite{liu2021characterizing,liao2023smartstate}, these mechanisms remain application-specific rather than properties of the settlement layer itself, creating a structural asymmetry relative to card systems.
\end{enumerate}

\subsection{Structural Comparison of Card and Stablecoin Systems}

This section compares card networks and stablecoin payment systems at the model and lifecycle levels, clarifying how payment functions are organized and executed across participants.

\smallskip
\noindent\textbf{Model Level Comparison.} At the model level, card networks and stablecoin systems differ fundamentally in how coordination, authority, and responsibility are organized among participants. Card networks implement a hierarchical, institutionally coordinated architecture in which issuers, acquirers, and schemes perform clearly delineated roles under contractual rulebooks. Authorization authority, risk management, and settlement obligations are centralized within regulated intermediaries, while schemes provide standardized governance, messaging, and interoperability across the network.

Stablecoin payment systems adopt a modular, protocol-coordinated architecture. Core payment functions are distributed across token issuers, blockchain networks, wallets, and intermediaries, with execution and settlement embedded in software rather than enforced through bilateral contractual arrangements. Stablecoin issuers are responsible for minting, burning, and reserve management but do not authorize individual retail transactions. Blockchain networks execute and finalize transfers according to consensus rules while remaining operational infrastructure rather than contractual counterparties. Intermediaries such as custodial platforms and exchanges aggregate liquidity, compliance, and access to fiat systems on a discretionary, market-driven basis rather than under scheme-mandated obligations.

Therefore, card networks concentrate governance and responsibility within a small number of institutional actors, whereas stablecoin systems unbundle these functions across multiple technical and organizational layers. This difference does not imply superiority of one model over the other, but it produces distinct patterns of authority, accountability, and risk allocation that shape how each system operates in practice.

\smallskip
\noindent\textbf{Lifecycle Level Comparison.} 
At the lifecycle level, the two systems diverge in how informational processing and value transfer are sequenced and finalized, as summarized in Table~\ref{tab:lifecycle-comparison}. In card networks, authorization, clearing, and settlement are temporally separated. Issuer authorization provides an immediate but conditional payment signal, while clearing and net settlement occur subsequently through scheme-defined processes and interbank settlement arrangements. Even after settlement, transactions remain subject to post-settlement recourse under scheme rules, preserving the possibility of reversal through standardized dispute mechanisms.

In stablecoin systems, transaction validation, ordering, and settlement are integrated within a unified ledger execution process. Once a transaction satisfies protocol-level admissibility and is confirmed under the network’s consensus mechanism, balances are reassigned directly on-chain. Clearing and settlement are therefore collapsed into a single state transition, and finality is determined by technical confirmation rather than institutional agreement. Native mechanisms for chargeback or reversal are absent at the base layer; any corrective action must be implemented through subsequent transactions or application-specific controls outside the settlement layer itself.

These lifecycle differences reflect alternative design choices rather than differences in transaction stages per se. Card systems preserve legal and economic conditionality beyond settlement, whereas stablecoin systems emphasize atomic execution and ledger finality. Assessing whether these design choices support substitutability in retail payments therefore requires an explicit, function-oriented evaluation framework.

\begin{table*}[htbp]
\footnotesize
\centering
\caption{Alignment of Transaction Lifecycle Stages Across Card and Stablecoin Payment Systems}
\label{tab:lifecycle-comparison}

\renewcommand{\arraystretch}{1.05}
\setlength{\tabcolsep}{4pt}

\begin{tabularx}{\linewidth}{@{}
    >{\centering\arraybackslash}m{2.8cm}
    >{\raggedright\arraybackslash}X
    >{\raggedright\arraybackslash}X
@{}}
\toprule
\makecell[c]{\bfseries Stage} & \textbf{Card Network Model} & \textbf{Stablecoin Model} \\
\midrule

\rowcolor[gray]{0.96}
\StageCell{4}{Initialization} &
\vspace{-1\baselineskip}
\begin{itemize}[leftmargin=*, nosep, topsep=0pt, partopsep=0pt]
  \item Capture payment credentials at the POI.
  \item Construct an authorization request under scheme messaging standards such as ISO 8583.
  \item Populate transaction parameters for subsequent network processing.
\end{itemize}
&
\vspace{-1\baselineskip}
\begin{itemize}[leftmargin=*, nosep, topsep=0pt, partopsep=0pt]
  \item Retrieve the merchant receiving address and construct a transfer instruction.
  \item Authorize the instruction through a cryptographic signature.
  \item Broadcast the signed transaction, or submit it via a custodial intermediary.
\end{itemize}
\\

\StageCell{4}{Authorization/Validation} &
\vspace{-1\baselineskip}
\begin{itemize}[leftmargin=*, nosep, topsep=0pt, partopsep=0pt]
  \item Validate the payer account and assess available funds or credit.
  \item Apply issuer fraud screening and risk controls.
  \item Approve or decline and place an authorization hold without moving interbank liquidity.
\end{itemize}
&
\vspace{-1\baselineskip}
\begin{itemize}[leftmargin=*, nosep, topsep=0pt, partopsep=0pt]
  \item Verify protocol admissibility, including signature correctness.
  \item Enforce transaction sequencing rules to prevent replay and preserve ordering.
  \item Confirm balance sufficiency and other protocol constraints prior to execution.
\end{itemize}
\\

\rowcolor[gray]{0.96}
\StageCell{4}{Clearing/Ordering} &
\vspace{-1\baselineskip}
\begin{itemize}[leftmargin=*, nosep, topsep=0pt, partopsep=0pt]
  \item Aggregate authorized transactions into clearing batches.
  \item Compute obligations across issuers and acquirers, often on a net basis.
  \item Translate transaction records into settlement positions under scheme rules.
\end{itemize}
&
\vspace{-1\baselineskip}
\begin{itemize}[leftmargin=*, nosep, topsep=0pt, partopsep=0pt]
   \item Continuously sequence admissible transactions under the network consensus mechanism.
   \item Commit the canonical execution order in blocks or execution batches.
   \item Determine state transitions directly from the committed transaction order.
\end{itemize}
\\

\StageCell{4}{Settlement} &
\vspace{-1\baselineskip}
\begin{itemize}[leftmargin=*, nosep, topsep=0pt, partopsep=0pt]
    \item Transfer net obligations after clearing, typically T+1 to T+2.
    \item Settle in interbank money, commonly via central bank infrastructures such as RTGS.
    \item Treat settlement finality as legally conditional under scheme rules and dispute frameworks.
\end{itemize}
&
\vspace{-1\baselineskip}
\begin{itemize}[leftmargin=*, nosep, topsep=0pt, partopsep=0pt]
  \item Update ledger state when a block is accepted under consensus.
  \item Reassign balances at the protocol level as an atomic state transition.
  \item Rely on the network finality model and merchant confirmation policy for effective acceptance.
\end{itemize}
\\

\rowcolor[gray]{0.96}
\StageCell{4}{Recourse} &
\vspace{-1\baselineskip}
\begin{itemize}[leftmargin=*, nosep, topsep=0pt, partopsep=0pt]
  \item Provide dispute handling under scheme rulebooks.
  \item Allow chargebacks and adjustments that can reverse prior outcomes.
  \item Reallocate liability contractually after the original transaction.
\end{itemize}
&
\vspace{-1\baselineskip}
\begin{itemize}[leftmargin=*, nosep, topsep=0pt, partopsep=0pt]
  \item Provide no native chargeback mechanism at the base ledger.
  \item Require corrective action through a subsequent transfer initiated by the recipient.
  \item Support escrow or dispute logic only through application-specific mechanisms.
\end{itemize}
\\

\bottomrule
\end{tabularx}
\end{table*}

\section{CLEAR: A Comparison Framework}
\label{sec:clear}

As illustrated previously, a fundamental structural divergence separates incumbent card systems from emerging stablecoin arrangements, spanning both the organization of authority at the model level and the sequencing of validation, settlement, and recourse at the lifecycle level. While these differences clarify how transactions are executed and governed, they are insufficient on their own to determine whether stablecoins can effectively substitute for cards in retail payments.

The substitutability question is inherently functional rather than architectural. Retail payment adoption is widely understood as a coordination problem involving multiple stakeholder groups, in which improving performance along a single dimension is insufficient if frictions persist elsewhere in the payment chain~\cite{rysman2009economics}. Regardless of how coordination and settlement are implemented, a retail payment instrument must therefore satisfy a common set of performance requirements across key economic, legal, operational, and adoption dimensions. To evaluate whether stablecoin-based payments meet these requirements relative to card networks, we introduce the \textbf{CLEAR} framework:

\begin{itemize}
    \item \textbf{C -- Cost.} Evaluates economic incentives for adoption. It compares the merchant-facing fee structures of card payment systems with the transaction cost dynamics of stablecoin-based payments, including protocol fees and network congestion effects.

\item \textbf{L -- Legality.} Evaluates the strength of the legal and regulatory safety net. It examines the trade-off between the ex post consumer protections embedded in card networks, such as chargebacks and dispute resolution, and the settlement finality characteristic of blockchain-based payments.

\item \textbf{E -- Experience.} Evaluates user-facing friction at the POI. It compares the speed, predictability, and usability of card-based interactions with the wallet-based transaction approval processes required in stablecoin payments.

\item \textbf{A -- Architecture.} Evaluates systemic and operational risk. It contrasts the centrally governed security and liability structures of card schemes with the distributed execution environments of stablecoin systems, where risk is fragmented across custodial, protocol, and infrastructure layers.

\item \textbf{R -- Reach.} Evaluates network effects and adoption constraints. It assesses the difficulty of coordinating merchants and users in a fragmented stablecoin ecosystem relative to the near-universal acceptance achieved by established card networks.
\end{itemize}

\subsection{Cost: The Inversion of the Pricing Structure}
\label{subsec:cost}

The divergence between card and stablecoin payment models is most economically visible in how transaction costs are allocated across market sides. Drawing on the theory of two-sided markets~\cite{rochet2002cooperation, rysman2009economics}, we argue that the key distinction lies in an inversion of the pricing structure: card networks subsidize the payer side while recovering costs from merchants, whereas stablecoin-based payments reduce certain merchant-facing charges at the rail level but introduce payer-facing marginal costs and operational frictions that weaken the subsidy mechanism underpinning modern retail payment habits. These differences generate contrasting patterns of cost incidence, pricing structure, and user-facing frictions.

\smallskip
\noindent\textbf{Merchant Acceptance Cost.} This dimension captures the marginal fee burden imposed on the merchant at the point of settlement. In card-based payment systems, the dominant merchant-facing charge is the MDR, expressed as a percentage of the transaction value deducted during settlement. MDR aggregates multiple underlying components, including interchange paid to the issuing bank, network or scheme fees, and acquirer or processor markups.

\textit{Card -- Merchant Discount Rate.} In the United States (US), the MDR, which comprises interchange, scheme, and processor fees~\cite{helcim2024}, typically ranges from about 1\% to 3\% of the transaction value in 2025~\cite{investopedia2025}. 
By contrast, under the European Union’s Interchange Fee Regulation (IFR), the interchange component of the MDR, which is the single largest fee element in card schemes, is capped at 0.2\% for consumer debit cards and 0.3\% for consumer credit cards~\cite{EUReg2015}.
The regulation substantially compresses average merchant acceptance costs relative to the US, although total MDR remains heterogeneous across merchant categories, card types, and acquirer contracts~\cite{ECB2010_PaymentSystem, FederalReserve2024}. 
Importantly, MDR is not merely a toll: it finances cross-subsidization within the two-sided platform. As formalized by Rochet and Tirole~\cite{rochet2002cooperation}, interchange allows issuers to subsidize consumer participation through rewards, grace periods, and fraud and liability services, helping the platform overcome the initial coordination problem of adoption. Merchants therefore pay a premium to access a broad base of consumers and the liquidity provided by issuer credit lines.

\textit{Stablecoin -- Disintermediated Transfer with Re-intermediation.} Stablecoin transfers remove interchange and scheme fee layers at the rail level by settling on a public ledger. In a direct transfer, the recipient-side fee is typically zero, while marginal on-chain costs are borne by the sender via gas fees. In practice, however, most retail merchants require custodial processors for compliance, key management, accounting, and often automated conversion to fiat. These services typically charge a percentage fee that functions as a de facto floor for crypto acceptance costs~\cite{BankCanada2022}. The resulting model resembles fee-for-service payment processing rather than fee-for-credit intermediation.

While stablecoins can therefore reduce certain acceptance fees in principle, they also weaken a key element of the card value proposition: the integration of consumer credit and its associated subsidy. In their baseline transfer form, stablecoins behave closer to digital cash than to digital credit, which may limit their ability to stimulate credit-fueled consumption absent an additional credit layer.

\smallskip
\noindent\textbf{Consumer Transaction Cost.}
\label{subsubsec:consumercost} 
This dimension evaluates the marginal cost and operational burden imposed on the payer at the point of transaction.

\textit{Card -- Near-zero Marginal Cost at POI.} In the card model, consumers typically face zero marginal fees at the point of interaction, while issuer competition for top-of-wallet status often takes the form of rewards and float (e.g., an interest-free grace period). In two-sided market terms, these features constitute a subsidy or negative price on the payer side, facilitating network take-up~\cite{rochet2002cooperation}. Consumer costs may still arise indirectly (e.g., through interest on revolving credit or price pass-through by merchants), but they are not typically levied as a per-transaction charge at POI.

\textit{Stablecoin -- Resource Metering and Asset Mismatch.} In blockchain-based payments, users (or their service providers) must fund state changes via network transaction fees (e.g., gas), which reflect the computational resources required to execute and record transfers~\cite{Buterin2014}. Unlike fixed banking fees, gas prices are determined in a market for block space and can vary with network conditions~\cite{donmez2022transaction}. Layer-2 (L2) scaling solutions can compress average fees materially by batching transactions on a Layer-1 chain~\cite{gudgeon2020sok,sguanci2021layer}, though fee levels remain contingent on network design and congestion~\cite{CoinLaw2025L2Fees}. A structural friction nonetheless remains in many settings: fees must often be paid in the network’s native asset (e.g., ETH), even when the transfer amount is denominated in a stablecoin. This fee-payment asset mismatch forces users to maintain balances in volatile native tokens to enable stablecoin payments, increasing operational complexity relative to single-asset card interactions~\cite{Li2024MStableChain}.

Thus, the cost dimension highlights an adoption-relevant reallocation: card networks externalize marginal frictions away from the payer at POI, whereas stablecoin payments tend to reintroduce payer-facing costs and wallet-management burdens unless fees are abstracted or subsidized by intermediaries.

\smallskip
\noindent\textbf{Liquidity and Boundary Cost.}
This dimension evaluates the cost of converting the settlement asset into working capital usable for routine fiat-denominated expenses.

\textit{Card -- Integrated Liquidity.} Card settlement ultimately delivers commercial bank money in the merchant’s domestic unit of account, enabling immediate use for taxes, wages, rent, and suppliers. For domestic transactions, conversion is typically not required; cross-border transactions can incur additional premia, often attributed to cross-region interchange schedules, network assessment fees, and processor currency conversion markups~\cite{EuropeanCommission2020IFR,VisaISA2023,StripePricing2025}.

\textit{Stablecoin -- Segregated Liquidity.} Stablecoin settlement delivers a tokenized asset distinct from the merchant’s fiat working-capital needs, creating a boundary that must be crossed unless the merchant operates within a crypto-native value chain. Off-ramping introduces renewed intermediation and typically involves (i) exchange or processor spreads/fees and (ii) banking frictions such as wire fees and settlement delays~\cite{circleNextChapter}. 

This comparison suggests a total cost of ownership (TCO)~\cite{ferrin2002total} perspective: lower rail-level acceptance costs may be offset by conversion and reconciliation costs when merchants ultimately require fiat liquidity. Consequently, stablecoins realize their clearest cost advantage in a closed-loop environment where incoming stablecoins can be redeployed to pay suppliers without repeatedly crossing the crypto-fiat boundary~\cite{BankCanada2022}.

\vspace{0.5\baselineskip}
\noindent
\fcolorbox{blue!40}{blue!5}{
\begin{minipage}{0.92\linewidth}
\setlength{\parskip}{0.6em}

\textbf{Cost Level Finding} 

\emph{The important cost distinction is not settlement efficiency but cost incidence.} 

Card networks subsidize consumers and recover costs from merchants, with interchange financing credit provision, fraud protection, and consumer guarantees. Stablecoin systems remove interchange and some merchant-side fees, but also remove the mechanism that funds these subsidies, shifting costs to users through transaction fees, liquidity fragmentation, and off-ramping. Consequently, stablecoins currently resemble digital cash rather than a retail credit instrument, performing well in closed loops but remaining disadvantaged in open-loop retail settings.
\end{minipage}}

\subsection{Legality: The Protection-Finality Trade-off}
\label{sec:legal}

This dimension evaluates how payment systems allocate liability and define the moment at which obligations are discharged. In line with the payments-economics literature~\cite{kahn2002economics,kahn2009pay}, these issues reflect a fundamental design tension: systems must balance reversibility, which protects users from fraud and error, with settlement finality that protects the wider system from credit and insolvency risk. Card networks and stablecoin arrangements resolve this trade-off through radically different institutional mechanisms. These differences result in contrasting allocations of liability, protection, and settlement finality.

\smallskip
\noindent\textbf{Consumer Protection and Recourse.}
This part analyzes the allocation of liability when a transaction is unauthorized, fraudulent, or disputed.

\textit{Card -- Statutory Protection.} In card systems, strong consumer protection is embedded in law and scheme rules. Under PSD2 in Europe and regulation in US, consumer liability for unauthorized transactions is strictly capped, and issuers are mandated to reimburse cardholders promptly. Chargebacks operationalize this protection. When a dispute is filed, the issuer must provisionally credit the payer and recover the amount from the merchant. In economic terms, the network mutualizes fraud risk by shifting the burden toward merchants, a mechanism consistent with platform theories of optimal liability allocation~\cite{rochet2002cooperation,neyer2017blockchain}.

\textit{Stablecoin -- Asset Safeguards and Transfer Risk.} The regulatory approach for stablecoins, exemplified by the EU’s MiCA regulation, focuses on the issuer rather than the transaction, protects the asset but not the transfer.
\begin{itemize}
    \item \emph{Asset Protection:} MiCA imposes strict requirements on issuers of EMTs and ARTs. It mandates that issuers maintain 1:1 reserves and grants holders a permanent right of redemption at par~\cite{micaReg}. This protects the user from issuer insolvency or de-pegging.
    \item \emph{Transfer Risk:} However, regulations generally do not mandate reimbursement for user errors or ``Authorized Push Payment" fraud. Because the user controls the private key and signs the instruction, the protocol treats the transaction as valid. There is no protocol-level administrator with a general statutory obligation or native technical capability to reverse an individual transfer once confirmed on the ledger.~\cite{zetzsche2021markets}.
\end{itemize}

Beyond the EU’s MiCA framework, the US, Singapore, and Hong Kong have each introduced stablecoin-specific regulatory regimes, reflecting a coordinated shift toward treating stablecoins as regulated payment instruments rather than speculative crypto-assets. Across these jurisdictions, regulation converges on full reserve backing, issuer licensing, and enforceable redemption rights. In the US, initiatives under the GENIUS Act emphasize reserve composition and supervisory oversight without introducing card-scheme-style dispute resolution mechanisms~\cite{US2025GENIUS}. Monetary Authority of Singapore (MAS) similarly prioritizes reserve backing and redemption rights for single-currency stablecoins while preserving on-chain transfer finality~\cite{mas2023stablecoin}, and Hong Kong’s proposed regime focuses on issuer authorization and governance rather than post-settlement consumer recourse~\cite{hkma2024sandbox}.

However, this convergence is largely issuer- and asset-centric rather than transaction-centric. While prevailing regimes ensure that stablecoins are prudently managed and redeemable at par, they provide limited safeguards once a payment has been executed. Transfers authorized at the wallet or protocol level are typically treated as final, even when induced by error or fraud, with no statutory obligation or native system mechanism for reversal. Unlike card networks, which mutualize fraud and dispute risk through issuer guarantees and scheme-level rules, stablecoin systems largely externalize transfer risk to end users, particularly in non-custodial settings, shifting the burden of security and error prevention from the payment system to the individual consumer. This structural distinction underpins the persistent legal and assurance gap between stablecoin-based payments and mature card-based retail payment systems.

\smallskip
\noindent\textbf{Settlement Finality and Enforceability.}
This dimension evaluates the legal moment when a payment obligation is irrevocably discharged.

\textit{Card -- Designated Finality.} Under frameworks such as the EU Settlement Finality Directive (SFD)~\cite{EUDirective}, payment systems designate a precise moment, usually the posting of entries at the central bank, after which transfers are legally irrevocable and protected from clawback, even in insolvency. This deterministic finality is foundational to modern clearing and settlement~\cite{cpmi2016glossary}.

\textit{Stablecoin -- Probabilistic Finality.} Stablecoins rely on blockchain consensus mechanisms that provide technical finality, including probabilistic finality in Proof-of-Work (PoW)~\cite{gervais2016security} systems and economic finality in Proof-of-Stake (PoS)~\cite{nguyen2019proof} systems, among others. While blocks become effectively irreversible after sufficient confirmations, this technical finality is not equivalent to legal finality. As noted by Nabilou~\cite{nabilou2020dark}, courts may recognize a payer’s claim even when the protocol cannot reverse the transfer, creating a mismatch between on-chain reality and off-chain legal obligations.

This dichotomy creates a strategic trade-off for merchants. Card network rails provide high legal certainty within established banking and legal frameworks, but expose merchants to post-settlement reversibility through consumer-initiated chargebacks. Stablecoin rails invert this balance: settlement is immediate and technically irreversible, offering strong operational certainty, yet the legal finality of payment remains ambiguous, particularly regarding whether the underlying obligation is definitively discharged under existing insolvency and commercial law.

\smallskip
\noindent\textbf{Data Privacy and Compliance.} A third legal friction exists regarding the compatibility of ledger architectures with data protection laws.

\textit{Card -- Private Ledgers.} Card network systems rely on private, centrally administered bank and network ledgers rather than publicly replicated state. Transaction data are held in controlled databases operated by issuers, acquirers, and card networks under contractual and regulatory governance. This centralized architecture is inherently compatible with the General Data Protection Regulation (GDPR), including the ``Right to be Forgotten"~\cite{GDPRArticle17}, as intermediaries can selectively delete or anonymize records upon valid erasure requests without affecting system operation, ensuring compliance through institutional governance rather than protocol-level design.

\textit{Stablecoin -- Public Verification.} Public permissionless blockchains~\cite{liang2024identity} function by broadcasting transaction details to a globally distributed network of nodes. Once data is written to the blockchain, it is technically impossible to erase. Even if the user’s identity is pseudonymous, represented only by a wallet address, metadata analysis can often re-identify individuals~\cite{biryukov2019deanonymization,xiang2023babd}. This immutability creates a direct tension with GDPR requirements for data erasure and rectification~\cite{finck2018blockchains}.

According to the above analysis, stablecoins face a compliance paradox. To be useful for retail payments, they require the transparency of a public ledger to ensure trust without an intermediary. Yet, this very transparency renders them potentially non-compliant with strict privacy regimes. Thus, widespread retail adoption may ultimately depend on the successful integration of privacy-preserving compliance mechanisms, including privacy-enhancing technologies such as Zero-Knowledge Proofs (ZKPs)~\cite{fiege1987zero,sun2021survey}, Secure Multiparty Computation (SMPC)~\cite{lindell2020secure,zhong2019secure}, and Trusted Execution Environments (TEEs)~\cite{jauernig2020trusted,lind2017teechain}, rather than on the base ledger design alone.

\vspace{0.5\baselineskip}
\noindent
\fcolorbox{blue!40}{blue!5}{
\begin{minipage}{0.92\linewidth}
\setlength{\parskip}{0.6em}

\textbf{Legality Level Finding}  
        
\emph{The decisive legal distinction in retail payments lies in whether protection is applied at the transaction level or confined to the asset level.} 

Card networks embed consumer protection directly into transaction processing through chargebacks, issuer guarantees, and scheme-level rulebooks that mutualize dispute risk across the system. Stablecoin regimes, by contrast, prioritize issuer solvency and redemption while preserving on-chain transfer finality. Without a network-wide rulebook or designated authority to adjudicate disputes, stablecoin payment systems remain institutionally incomplete, externalizing transfer risk to users and turning technical finality into a material business risk in open-loop retail payments.
\end{minipage}}

\subsection{Experience: The Perception-Performance Gap}\label{sec:experience}
The experience dimension synthesizes subjective perceptions such as trust, usability, and perceived security with objective system performance such as latency and reliability. Literature rooted in the Technology Acceptance Model (TAM)~\cite{marangunic2015technology} and the Unified Theory of Acceptance and Use of Technology (UTAUT)~\cite{venkatesh2016unified} consistently finds that adoption hinges less on technical capability and more on perceived usefulness, perceived security, and ease of use, which are constructs that shape both intention and actual usage~\cite{quan2023mobile}. These perceptual factors serve as economic complements to institutional assurances such as liability protection and settlement finality. These differences give rise to contrasting perceptions of security, usability, and reliability.

\smallskip
\noindent\textbf{Consumer Perception and the Demand Side.}
This part evaluates the psychological and structural barriers facing the payer.

\textit{Card -- Structural Assurance.} Consumers favor payment methods perceived as secure, simple, and useful. In mature markets, card networks benefit from what Sangari and Mashatan~\cite{sangari2024driving} define as tradition-related barriers regarding challengers: the user's preference for familiar methods is reinforced by deep structural assurances such as clear governance, defined recourse, and reliable intermediation. These assurances lower the threshold for trust, allowing intention to translate directly into use without significant cognitive resistance.

\textit{Stablecoin -- Consumer Resistance.} Barriers to adoption remain substantial. Although awareness of stablecoins is high, uptake is constrained by consumer resistance. 
A multi-analytical study by Sangari and Mashatan~\cite{sangari2024driving} shows that price volatility is not the sole impediment; rather, it intensifies underlying concerns related to the absence of structural assurances, including unclear protections and opaque governance. 
User-centered security studies further clarify the experiential mechanisms behind this resistance: while users associate stablecoins with price stability and perceived regulatory alignment, they remain uncertain about reserve transparency, governance, and failure modes across different stablecoin designs, and must actively perform risk assessment and mitigation in the absence of institutional guarantees~\cite{si2024understanding,guan2025security}. 
In the absence of such assurances, security and privacy concerns increase, reinforcing consumers’ reliance on established payment methods.

This resistance is reflected in a persistent gap between awareness and use. Survey evidence shows that awareness can be widespread, yet ownership and transactional use remain limited. For example, Canadian survey evidence reports that roughly 64\% of respondents were aware of Bitcoin in 2017, while only about 2.9\% reported owning it\cite{henry2018bitcoin}, and often motivated by investment rather than payment needs \cite{schuh2016us, shahzad2018empirical}.

Evidence from the hospitality sector further underscores this pattern. Quan and coauthors \cite{quan2023mobile} find that perceived security and ease of use are the dominant determinants of trust. The operational complexity of managing private keys in self-custodial stablecoin systems fails to meet the ease-of-use threshold required to displace mobile wallets in high-frequency retail environments.

The broader literature consistently shows that trust mediates the relationship between technology and user intention. Stablecoins currently exhibit a trust deficit: although they provide functional advantages such as speed and global reach, they lack the institutional safety wrap that consumers depend on to mitigate risk.

\smallskip
\noindent\textbf{Merchant Perception and the Supply Side.}
This section evaluates the willingness of retailers to integrate new payment rails.

\textit{Card -- Network Externalities.} Merchant acceptance is driven by expected net transactional benefits and observable customer demand~\cite{jonker2019drives}. Card networks benefit from strong positive network externalities: because nearly all consumers carry cards, merchants are compelled to accept them to avoid losing sales.

\textit{Stablecoin -- The Niche Trap.} Survey evidence indicates that crypto acceptance is not random but highly conditional. Polasik et al.~\cite{polasik2015price} find that adoption is most visible among start-ups, smaller retailers, and firms operating in developing economies or informal sectors, where weak formal infrastructure makes crypto a rational substitute for legacy bank and card payment rails. For mainstream retailers, by contrast, the primary barrier is the absence of customer demand \cite{jonker2019drives}. This creates a feedback loop in which consumers do not pay with crypto because merchants do not accept it, and merchants do not accept it because consumers do not request it. Silinskyte \cite{silinskyte2014understanding} further notes that facilitating conditions, particularly the ease of integrating through payment service providers, are essential for translating merchant intention into actual implementation.

This evidence points to a classic two-sided market coordination failure. Unlike the card network, which utilized interchange fees to subsidize the chicken-and-egg problem, the stablecoin ecosystem lacks a centralized mechanism to simultaneously incentivize merchants and consumers. Consequently, uptake remains modest and fragmented outside of specific niches.

\smallskip
\noindent\textbf{Operational Performance.}
\label{operation}
This section remains focused on the objective metrics about latency reliability we discussed previously: approved and confirmation. Consequently, even where technical performance converges at the user interface level, differences in predictability, failure handling, and institutional backstops sustain a perceptual gap between card-based and stablecoin-based payments.

\textit{Card -- Instant Operational Approval.} 
Card authorization is commonly described as completing within a few seconds, which merchants interpret as a near real-time operational signal that settlement will follow~\cite{checkout2024_cardAuthorization}.

\textit{Stablecoin -- Routing and Liquidity.} While some blockchain networks and scaling solutions can deliver near real-time user-facing payment acknowledgments, performance remains sensitive to network conditions, routing pathways, and available liquidity. Providers typically display a paid or pending status within seconds, yet final confirmation may take longer, particularly on base-layer networks where settlement depends on block production times~\cite{malakhov2023analysis,kawase2020priority}. A further limitation is the lack of audited, cross-rail performance statistics; existing comparisons rely largely on heterogeneous vendor reports rather than harmonized measurement frameworks.

\vspace{0.5\baselineskip}
\noindent
\fcolorbox{blue!40}{blue!5}{
\begin{minipage}{0.92\linewidth}
\setlength{\parskip}{0.6em}

\textbf{Experience Level Finding}  

\emph{The critical determinant of user experience in retail payments is not raw technical performance but perceived security, predictability, and ease of use.} 

Card networks align system behavior with user expectations by embedding institutional guarantees and familiar interaction patterns that abstract risk away from users. Stablecoin payments, by contrast, shift security and error management to consumers and merchants, increasing cognitive and operational burden even when underlying performance is comparable. Thus, user experience reflects how effectively a payment system minimizes uncertainty at the point of interaction.
\end{minipage}}

\subsection{Architecture: System Integrity and Security}
\label{sec:architecture} 

The architectural dimension concerns how payment systems validate transactions, maintain ledger integrity, and withstand operational failures or attacks. In payments economics, these choices are shaped by the underlying instrument type. Account-based systems require a trusted authority to verify identity and update balances, whereas token-based systems require distributed consensus to validate transfers of bearer-like assets \cite{kahn2009pay}. These differences shape contrasting patterns of operational resilience and security risk allocation.

\smallskip
\noindent\textbf{Network Topology and Operational Resilience.} This section focuses on how structural layout affects system resilience and the pattern of failures observed in practice.

\textit{Card -- Centralized Intermediation.}
Traditional card networks follow a hub-and-spoke topology in which a central intermediary coordinates messaging, risk scoring, and routing between issuers and acquirers. This structure optimizes throughput and operational control, enabling near-instant authorization and globally interoperable acceptance. However, literature in financial market infrastructures notes that centralization introduces correlated-failure risks: outages at a major processor, scheme switch, or cloud provider can halt retail transactions at scale \cite{moore2010economics,cpmi2016glossary}. Empirically, card outages in Europe and the US illustrate this fragility, yet systematic academic quantification of such events remains limited.

\textit{Stablecoin -- Distributed Consensus with Centralized Bottlenecks.}
Stablecoin arrangements inherit the security and fault-tolerance properties of the blockchains~\cite{gramoli2024stabl} on which they are issued. These networks implement forms of Byzantine Fault Tolerant (BFT)~\cite{castro1999practical,sousa2018byzantine} consensus that allow decentralized validators to maintain ledger integrity under adversarial conditions. In blockchain-based payment environments, this produces a settlement layer that is resilient to validator outages and capable of maintaining continuity of the ledger \cite{nakamoto2008bitcoin,bohme2015bitcoin}.

At the user-facing level, however, operational fragilities arise from reliance on intermediated transaction processing. Even when settlement is decentralized, many practical implementations depend on a single operator or service to coordinate transaction flow before it reaches the underlying blockchain~\cite{motepalli2023sok}. This creates a structural asymmetry: the base ledger exhibits strong theoretical resilience, yet the execution environment can stall if the coordinating service experiences downtime, congestion, or restrictive behavior. Although users may retain the ability to move their balances directly on the base chain, they cannot access the continuous, low-latency transfers expected in retail commerce when such interruptions occur. Empirical research highlights this gap between settlement robustness and operational fragility as a key bottleneck for stablecoin use in payments \cite{thibault2022blockchain,mccorry2021sok}.

\smallskip
\noindent\textbf{Security Model Shift.} \label{subsubsec:securitymodel} This section analyzes how architectures allocate security risk, contrasting institution-managed credential protection with user-managed cryptographic control.

\textit{Card -- Credential-sharing and Perimeter Security.}
Account-based networks rely on credential sharing, including Primary Account Number (PAN) and other personal identifiers, which must be processed across many intermediaries. This places concentrated volumes of sensitive information in large custodial databases. Security economists describe these repositories as honeypots that attract attackers and cause repeated large-scale breaches~\cite{anderson2010security}. Although tokenization and Payment Card Industry Data Security Standards (PCI DSS)~\cite{morse2008pci} reduce some exposure, the overall perimeter-defense model still depends on institutional controls rather than mathematical certainty.

\textit{Stablecoin -- Cryptographic Guarantees and User Vulnerability.}
In token-based architectures, security is achieved via cryptography rather than identity verification~\cite{nakamoto2008bitcoin}. The private key acts as the authorization root and is never shared with the merchant, so the system structurally eliminates the institutional honeypot risk inherent in card networks (where centralized databases of PANs attract mass breaches). However, this architecture creates a security inversion, shifting the locus of risk from the network perimeter to the user endpoint.

Empirical studies emphasize that while the network is resilient, the user is fragile. Karantias et al.~\cite{karantias2020sok} document the high frequency of wallet compromise, phishing, and key loss-failures for which blockchain protocols offer no formal remediation mechanism. Unlike the card model, where the issuer absorbs the risk of fraud, the stablecoin model imposes strict liability on the user. Furthermore, the reliance on programmable settlement introduces smart contract risk, where software exploits create a vector for systemic loss absent in traditional banking ledgers~\cite{he2020smart,zhuang2021smart,chen2025chatgpt}. Recent research highlights these technical risks but frequently overlooks the economic implication: that stablecoins effectively demonetize the insurance function of payments, trading consumer protection for settlement finality.

\vspace{0.5\baselineskip}
\noindent
\fcolorbox{blue!40}{blue!5}{
\begin{minipage}{0.92\linewidth}
\setlength{\parskip}{0.6em}

\textbf{Architecture Level Finding}  
            
\emph{The fundamental architectural trade-off in retail payments lies between centralized risk absorption and decentralized technical resilience.} 

Card networks centralize validation and security within intermediaries, enabling strong institutional control but exposing the system to correlated operational failures. Stablecoin systems, by contrast, secure ledger integrity through distributed consensus and cryptographic verification, reducing settlement risk while shifting execution and security risk to users and smart contracts. Therefore, architecture determines not only system resilience but also where operational and security risks ultimately reside within the payment ecosystem.
\end{minipage}}

\subsection{Reach: The Fragmentation Barrier}
\label{sec:reach}

The reach dimension concerns the scale of merchant acceptance and the interoperability standards that enable it. Payment systems are classic network goods: their value rises with the size of their user base, and incumbents benefit from strong positive feedback loops, switching costs, and path dependence \cite{katz1985network,farrell1986installed}. These dynamics shape adoption more powerfully than technical performance alone. In this sense, differences in reach reflect not only technical design choices but also the presence or absence of mechanisms that coordinate expectations across market participants. These architectural differences manifest along two closely related dimensions: network topology and operational resilience, and the underlying security model governing transaction authorization and risk allocation. These differences translate into contrasting degrees of interoperability, coordination, and network scale.

\smallskip
\noindent\textbf{Network Scale and Standardization.} This section compares how card networks and stablecoin systems achieve reach, highlighting the contrasting roles of standardization and fragmentation in shaping their global scalability and reliability.

\textit{Card -- Global Interoperability.}
Traditional card networks achieve near-universal reach through decades of enforced standardization. Interoperability is secured through ISO 8583~\cite{ISO_8583_2023} messaging, EMV hardware specifications and scheme rulebooks that ensure functional equivalence across issuers, acquirers and geographies. With an acceptance footprint now exceeding 150 million merchant locations in more than 200 countries and territories~\cite{nilsonreport2024}, cards exhibit the characteristics of a mature network good: a ubiquitous installed base and stable expectations that reinforce adoption on both sides of the market. The two-sided platform literature emphasizes that such standardization is essential for global scale~\cite{evans2004paying}. Taken together, these features give card networks a highly predictable operational profile and make them difficult to fully displace even when alternative technologies emerge.

\textit{Stablecoins -- Fragmented Infrastructures.}
Stablecoin networks operate across heterogeneous blockchains, such as Ethereum, Solana, and Tron~\cite{li2025blockchain1}, each with distinct execution environments and token standards. This creates structural fragmentation, since identical stablecoins issued across chains are not mutually substitutable and no shared messaging layer exists to coordinate them. This fragmentation is not merely technical but reflects a coordination failure, as no governing entity exists to mandate standards or align expectations across issuers, wallets, and merchants. Interoperability depends on cross-chain bridges~\cite{xie2022zkbridge,lee2023sok} that lock assets on one chain and mint representations on another. Empirical data shows that cross-chain bridges were a particularly vulnerable component of the blockchain ecosystem, with an estimated \$2 billion stolen across 13 separate bridge hacks—accounting for a majority of total value lost in crypto hacks in 2022 ~\cite{chainalysis2023}. Although technical proposals aim to improve portability and cross-chain execution, most remain early-stage and are unevenly implemented. As a result, stablecoin circulation remains dispersed across isolated execution environments, and operational behavior varies widely depending on chain-level design choices.

\smallskip
\noindent\textbf{Coordination Problem.} This section examines how coordination dynamics shape adoption, contrasting the strong network effects that reinforce card dominance with the more fragile and localized patterns seen in stablecoin use.

\textit{Card -- Lock-in via Network Effects.}
Large installed bases generate substantial switching costs. A merchant who drops cards risks losing nearly all customers, and a consumer who abandons cards faces limited acceptance for everyday purchases. This dynamic reflects the start-up problem in network economics, that is, even if a challenger technology is technically superior, it cannot displace an incumbent without overcoming entrenched expectations and coordination equilibria~\cite{katz1985network}. Empirical work on payment-system competition shows that merchants adopt new methods only when expected demand is sufficiently large or when intermediaries provide financial incentives to offset the risk of early adoption. These reinforcing expectations help maintain the stability of card-based retail payments across diverse markets.

\textit{Stablecoin -- Limited to Closed Loops.}
Stablecoin payments, by contrast, tend to flourish in closed-loop environments such as crypto exchanges, remittance corridors and merchant niches in economies with weak banking infrastructure. In advanced economies, the lack of unified standards and the absence of broad consumer demand limit the emergence of mutually reinforcing network effects. As stablecoin activity remains distributed across isolated platforms and user communities, adoption patterns are driven more by localized incentives and specific operational contexts than by broad two-sided market forces. This leads to a payment landscape where stablecoins function within domain-specific ecosystems but exhibit limited capacity to generalize into mass retail environments.

Card networks operate as a unified system because interoperability is mandated and consistently enforced across all participants. Stablecoins, by contrast, function across a set of incompatible ledgers that lack a common settlement or messaging layer, which keeps circulation dispersed and coordination limited. Without convergence on shared technical and operational standards similar to ISO 8583 or EMV, stablecoin arrangements are unlikely to achieve the broad, retail-facing reach characteristic of card networks. In brief, these structural differences explain why cards scale as a global payment utility while stablecoin usage remains concentrated within specific, self-contained environments.

\vspace{0.5\baselineskip}
\noindent
\fcolorbox{blue!40}{blue!5}{
\begin{minipage}{0.92\linewidth}
\setlength{\parskip}{0.6em}

\textbf{Reach Level Finding}

\emph{The primary determinant of reach in retail payments is not technological openness but coordinated standardization and expectation alignment.}

Card networks achieve global scale by enforcing interoperable standards that consolidate participants into a single, predictable system and enable strong network effects across merchants and consumers. Stablecoin payments, by contrast, fragment activity across heterogeneous ledgers and interoperability layers, dispersing liquidity and usage norms. As a result, stablecoins struggle to generate self-reinforcing network effects and remain largely confined to closed or domain-specific loops in the absence of system-wide coordination.
\end{minipage}}

\section{Discussion and Implications}
\label{sec:discussion}

The preceding sections evaluated stablecoin-based payments relative to incumbent card networks along the CLEAR dimensions. This discussion interprets those results in a broader economic framework, focusing on the conditions under which stablecoins outperform incumbent rails, the constraints on convergence in retail payments, and the implications for competition and coexistence with central bank digital currencies (CBDCs).

\subsection{Contextual Domains of Stablecoin Advantage}

Stablecoin-based payment arrangements demonstrate strong technical capabilities at the settlement layer, yet remain institutionally incomplete as general-purpose retail payment instruments. Whether convergence with incumbent card networks is feasible depends not on eliminating all frictions, but on selectively mitigating those that directly constrain adoption. This subsection interprets the preceding analysis by identifying the conditions under which stablecoins may partially converge with card-based systems, while remaining structurally distinct.

\smallskip
\noindent\textbf{High-friction Cross-border Corridors.}
Stablecoins deliver their most pronounced efficiency gains in cross border remittances and Business-to-Business (B2B) treasury flows, where legacy correspondent banking chains remain both costly and slow. Empirical evidence shows that the global average cost of sending the benchmark remittance amount of \$200 reached 6.35\% in Q1 2024, while banks charged an average of 12.66\%, substantially exceeding the fees of money transfer operators and digital providers~\cite{worldbank2024remittance,ffnews2024}. These costs reflect the layered structure of correspondent banking, which relies on sequential intermediaries, manual reconciliation, and limited operating hours.

In this context, stablecoin based settlement compresses both time and cost by bypassing multiple correspondent layers and enabling near continuous ledger based finality. High throughput networks support rapid cross border value transfer with fewer intermediaries, making efficiency gains particularly salient where traditional cross-border payment infrastructure performs poorly. Industry estimates indicate that in 2024, approximately 5-10 \% of stablecoin transaction volume, equivalent to around 1.3 trillion US dollars, reflected genuine real economy payment activity, including cross border remittances, corporate treasury operations, and retail payments, with comparative advantages concentrated in cross border use cases~\cite{jhanji2025stablecoins}.

\smallskip
\noindent\textbf{High-Inflation and Financially Constrained Economies.}
A second domain of stablecoin advantage arises in jurisdictions experiencing sustained inflation, currency depreciation, or limited access to reliable banking services. In these settings, the primary appeal of stablecoins lies less in payment efficiency than in access to a relatively stable unit of account and store of value. USD denominated stablecoins in particular function as de facto instruments of value preservation in markets such as Turkey and Nigeria, where domestic currencies exhibit pronounced volatility~\cite{jhanji2025stablecoins}.

Here, features that are often viewed as constraints in advanced retail payment systems, most notably the immutability of blockchain ledgers, take on a different functional role. Ledger immutability provides users with a form of monetary autonomy that allows savings to be insulated from domestic banking instability, capital controls, or inflationary erosion. As a consequence, stablecoin usage in these economies reflects a demand for monetary stability rather than a substitution for mature retail payment instruments. In this sense, stablecoins operate as an alternative monetary infrastructure rather than as a competing retail payment rail.

\subsection{Pathways to Partial Convergence}

Stablecoin arrangements demonstrate strong technological performance in settlement through atomic execution, yet remain institutionally incomplete as general-purpose retail instruments because they lack a scheme-level governance rulebook. The shift from hierarchical, institution-managed networks to modular, protocol-governed architectures creates a responsibility vacuum: dispute handling, liability allocation, and operating standards are pushed to wallets, intermediaries, and courts, which weakens consumer recourse and inhibits the self-reinforcing network effects seen in card schemes. 

Accordingly, convergence depends less on raw throughput and more on developing governed intervention mechanisms and shared operational standards that can replicate the trust and stability of incumbent systems. This subsection synthesizes the prior evidence into practical pathways toward partial convergence.

\smallskip
\noindent\textbf{Reconstructing Consumer Side Subsidies.}
Convergence in retail payments is shaped by whether stablecoin arrangements can recreate the consumer-side subsidies that historically enabled card network adoption. While stablecoin settlement compresses merchant-facing acceptance costs, widespread retail use requires that transaction-related frictions be abstracted away from consumers. In their current form, gas fees and portfolio management requirements impose operational burdens on users, which conflicts with the pricing structure that underpins habitual card usage.

Progress toward convergence therefore depends on reallocating these burdens to intermediaries that benefit from stablecoin-based settlement. Account abstraction under ERC 4337 enables paymasters to sponsor transaction fees and allows payment service providers to internalize costs within their treasury operations~\cite{buterin2021erc4337}. Early pilots by firms such as Worldpay and Nuvei illustrate this direction, as stablecoin-related costs are absorbed at the infrastructure level rather than exposed at the POI~\cite{worldpay2022usdc,nuvei2024crypto}. Such fee internalization mirrors card economics, where operational complexity is deliberately hidden from end users in order to facilitate scale.

\smallskip
\noindent\textbf{Reconciling Finality with Legal Recourse.}
Legal convergence is constrained by the tension between technical irreversibility and the demand for transaction-level protection. Stablecoin arrangements provide asset-level safeguards through reserve backing and redemption rights, yet generally lack mechanisms for consumer reimbursement in cases of fraud or mistaken transfers. Card-based systems, by contrast, embed chargebacks, liability allocation, and statutory reimbursement obligations that establish predictable outcomes when transactions fail and anchor consumer trust.

Movement toward convergence is emerging through hybrid regulatory and institutional designs. Frameworks such as MiCA introduce issuance controls and allow issuer- or authority-initiated freezes that permit limited intervention without replicating full card-style reversibility~\cite{micaReg}. At the same time, institutional actors such as Visa and PayPal demonstrate how compliance screening, fraud controls, and controlled intervention can be layered onto stablecoin-enabled payment environments. Over time, convergence requires legal doctrines that align technical finality with legally recognized discharge of obligations, producing predictability comparable to regimes such as PSD2~\cite{kahn2009pay}. These developments point toward partially governed architectures rather than a return to fully immutable instruments.

\smallskip
\noindent\textbf{Abstracting Risk and Interaction Frictions.}
Experiential convergence depends on narrowing the gap between system performance and user expectations at the POI. In mature retail markets, consumers expect payment experiences that are fast, predictable, and supported by strong structural assurances. Direct self-custodial use of stablecoins exposes users to key management, fee selection, and routing uncertainty, which increases perceived risk and suppresses willingness to transact. Empirical evidence consistently shows that trust and ease of use dominate payment choice, and stablecoin-based interactions remain disadvantaged on both dimensions~\cite{sangari2024driving,quan2023mobile}.

Mitigation therefore hinges on abstraction at the interface level. Account abstraction under ERC 4337~\cite{buterin2021erc4337} allows wallets to conceal gas fees, streamline authorization flows, and introduce familiar authentication mechanisms such as biometric verification or social recovery. Large platforms increasingly embed stablecoins within closed-loop environments, as illustrated by PayPal’s PYUSD, where identity verification and fraud controls remain intact~\cite{paypal2023pyusd}. In these configurations, users experience stablecoin payments as simple transactional actions rather than cryptographic procedures, bringing interaction patterns closer to those of card-based systems.

\smallskip
\noindent\textbf{Consolidating Operational Reliability.}
Operational convergence requires aligning resilient settlement with dependable end-to-end execution. Blockchain consensus mechanisms provide strong guarantees of ledger integrity, yet retail payment performance depends on the reliability of surrounding layers, including wallets, routing services, bridges, and liquidity providers. While stablecoin settlement is relatively robust by design, user experience can degrade when these ancillary components encounter congestion, downtime, or coordination failures~\cite{thibault2022blockchain,mccorry2021sok}.

Progress toward convergence therefore depends on architectural consolidation and institutional standardization. Industry pilots illustrate early movement in this direction. Visa’s USDC-based settlement integrates stablecoins directly into acquirer treasury flows, reducing reliance on legacy correspondent banking infrastructure~\cite{visa2023stablecoin}. Cross-chain interoperability initiatives such as the Cross-Chain Interoperability Protocol (CCIP) seek to standardize messaging and settlement across heterogeneous chains~\cite{chainlink2021ccip}, although most remain at an early stage and often rely on trusted components~\cite{belchior2021survey}. Achieving card-level reliability requires clearly defined operational rules, recovery procedures, and governance structures that ensure predictable behavior under stress.

\smallskip
\noindent\textbf{Coordinating Standards and Expectations.}
Scalable convergence ultimately depends on coordinating expectations across market participants through shared standards. Card networks achieve global reach by enforcing interoperability through ISO 8583 messaging, EMV authentication, and scheme rulebooks, which ensure consistent behavior across merchants, issuers, and geographies~\cite{ISO_8583_2023}. This uniformity underpins strong network effects and stabilizes adoption on both market sides.

Stablecoin ecosystems, by contrast, remain distributed across incompatible execution environments, with interoperability often mediated by bridges that introduce operational and security risks. Meaningful convergence therefore requires both technical and institutional coordination. Technically, stablecoins require shared settlement and messaging standards that deliver consistent functionality across chains. Institutionally, they require governance frameworks that define acceptable behavior for issuers, payment service providers, and intermediaries, allowing merchants to treat stablecoin payments as predictable and uniform. Until such coordination emerges, stablecoins are likely to remain highly effective within specialized domains, yet constrained in their ability to scale as a universal retail payment utility.

\subsection{Stablecoins and CBDCs in Systemic Competition}

Private stablecoins face inherent limitations in supplying the legal finality, supervisory oversight, and systemic safeguards required for large-scale retail deployment~\cite{bis2023blueprint,CPMI-IOSCO-2022-Stablecoin}. These constraints have prompted central banks to explore central bank digital currencies (CBDCs) not merely as technological upgrades to existing payment infrastructures, but as strategic instruments for preserving monetary sovereignty and anchoring trust in an increasingly tokenized financial system~\cite{auer2022central,prodan2024rise,bis2024tokenisation}. Recent policy blueprints converge on a layered architecture in which CBDCs operate as a foundational settlement asset, while private digital instruments provide higher-layer functionality and application-specific innovation.

Within this emerging architecture, competition between stablecoins and CBDCs operates at the level of system design rather than direct substitution. CBDCs are structurally advantaged in providing legally recognized finality, uniform compliance with consumer protection regimes, and native integration with anti-money laundering and supervisory frameworks~\cite{al2023anti}. They can also be embedded within domestic instant payment infrastructures and designed with negligible marginal transaction costs, reducing the scope for private instruments to compete on cost or baseline accessibility alone. From an architectural perspective, state-issued systems can enforce common interoperability standards across intermediaries, avoiding the fragmentation and bridge-related vulnerabilities that characterize multi-chain stablecoin ecosystems~\cite{CPMI-IOSCO-2022-Stablecoin}. These features do not ensure superior adoption outcomes, but they do confer a level of institutional credibility that private issuers cannot replicate independently.

The more plausible trajectory is therefore one of functional differentiation rather than wholesale displacement. Stablecoins retain comparative advantages in environments where programmability, composability, and cross-border portability are central, including platform-based economies, decentralized financial applications, and selected international payment corridors where sovereign systems may lack interoperability. CBDCs, by contrast, are positioned to serve as settlement anchors, supplying legal certainty, operational uniformity, and state-backed assurances suited to large-scale retail and institutional use. As sovereign digital instruments begin to replicate many of the technical features associated with stablecoins while closing institutional gaps around governance and recourse, competitive dynamics shift away from direct rivalry toward complementary roles. 

\section{Conclusion}
\label{sec:conclusion}

Stablecoins and card networks embody fundamentally different institutional and technical logics. Stablecoins deliver continuous, programmable settlement on shared ledgers but externalize core elements of the retail payment safety wrapper, including consumer recourse, liability allocation, and interoperability, into wallets, intermediaries, and courts. Card networks, while slower at final settlement, function more effectively as mass-market instruments by bundling authorization, risk management, and post-transaction protections into enforceable rulebooks. Consequently, stablecoins perform well in closed-loop or high-friction environments but remain structurally disadvantaged in open-loop POS settings where trust, simplicity, and near-universal acceptance dominate adoption.

Across our research questions, card networks emerge as pull-based, intermediary-governed account systems with separated authorization and settlement, whereas stablecoins operate as push-based token transfers that collapse validation and settlement into irreversible ledger updates (RQ1). Under the CLEAR framework (RQ2), stablecoins display a conditional cost advantage through 24/7 settlement and reduced rail-level fees, but underperform on legality, experience, and reach. Closing this gap (RQ3) requires selective re-intermediation, such as fee abstraction, governed intervention mechanisms, and shared standards, rather than full disintermediation. Absent such convergence, stablecoins are likely to complement card networks as efficient settlement assets and niche payment instruments, rather than replace them in general-purpose retail payments.

\bibliographystyle{IEEEtran}
\bibliography{refs}

\begin{thebibliography}{100}
\providecommand{\url}[1]{#1}
\csname url@samestyle\endcsname
\providecommand{\newblock}{\relax}
\providecommand{\bibinfo}[2]{#2}
\providecommand{\BIBentrySTDinterwordspacing}{\spaceskip=0pt\relax}
\providecommand{\BIBentryALTinterwordstretchfactor}{4}
\providecommand{\BIBentryALTinterwordspacing}{\spaceskip=\fontdimen2\font plus
\BIBentryALTinterwordstretchfactor\fontdimen3\font minus \fontdimen4\font\relax}
\providecommand{\BIBforeignlanguage}[2]{{%
\expandafter\ifx\csname l@#1\endcsname\relax
\typeout{** WARNING: IEEEtran.bst: No hyphenation pattern has been}%
\typeout{** loaded for the language `#1'. Using the pattern for}%
\typeout{** the default language instead.}%
\else
\language=\csname l@#1\endcsname
\fi
#2}}
\providecommand{\BIBdecl}{\relax}
\BIBdecl

\bibitem{BIS2003}
\BIBentryALTinterwordspacing
{Bank for International Settlements}, ``The role of central bank money in payment systems,'' Tech. Rep., 2003. [Online]. Available: \url{https://www.bis.org/cpmi/publ/d55.pdf}
\BIBentrySTDinterwordspacing

\bibitem{hasan2012retail}
I.~Hasan, T.~De~Renzis, and H.~Schmiedel, \emph{Retail payments and economic growth}.\hskip 1em plus 0.5em minus 0.4em\relax Bank of Finland Research Discussion Papers, 2012, no. 19/2012.

\bibitem{ECB2010_PaymentSystem}
\BIBentryALTinterwordspacing
{European Central Bank}, ``The payment system: Payments, securities and derivatives, and the role of the eurosystem,'' 2010, accessed: Oct. 10, 2025. [Online]. Available: \url{https://www.ecb.europa.eu/pub/pdf/other/paymentsystem201009en.pdf}
\BIBentrySTDinterwordspacing

\bibitem{hasan2013retail}
I.~Hasan, T.~De~Renzis, and H.~Schmiedel, ``Retail payments and the real economy,'' 2013.

\bibitem{lagator2021global}
A.~Lagator, ``Global development trends in payment card industry,'' National Bank of Serbia, Tech. Rep., 2021.

\bibitem{BIS2016}
\BIBentryALTinterwordspacing
{Bank for International Settlements}, ``Fast payments - enhancing the speed and availability of retail payments,'' 2016, accessed: Oct. 10, 2025. [Online]. Available: \url{https://www.bis.org/cpmi/publ/d154.pdf}
\BIBentrySTDinterwordspacing

\bibitem{ECB2020}
\BIBentryALTinterwordspacing
{European Central Bank}, ``Report on a digital euro,'' 2020, accessed: Oct. 10, 2025. [Online]. Available: \url{https://www.ecb.europa.eu/pub/pdf/other/Report_on_a_digital_euro~4d7268b458.en.pdf}
\BIBentrySTDinterwordspacing

\bibitem{yawe2020impact}
U.~B.~L. Yawe and U.~I. Mukisa, ``The impact of the revised payment services directive on the market for payment initiation services.'' \emph{Journal of Payments Strategy \& Systems}, vol.~14, no.~1, 2020.

\bibitem{defillama2025}
\BIBentryALTinterwordspacing
{DefiLlama}, ``Total stablecoin market capitalization,'' 2025, accessed: Nov. 23, 2025. [Online]. Available: \url{https://defillama.com/stablecoins}
\BIBentrySTDinterwordspacing

\bibitem{Rise2025Stats}
\BIBentryALTinterwordspacing
Rise, ``25 stablecoin statistics from 2025 (so far),'' 2025, accessed: Nov. 19, 2025. [Online]. Available: \url{https://www.riseworks.io/blog/stablecoin-statistics-from-2025}
\BIBentrySTDinterwordspacing

\bibitem{wang2022exploring}
Q.~Wang, R.~Li, Q.~Wang, S.~Chen, M.~Ryan, and T.~Hardjono, ``Exploring web3 from the view of blockchain,'' \emph{arXiv preprint arXiv:2206.08821}, 2022.

\bibitem{okx2025pay}
\BIBentryALTinterwordspacing
{OKX}, ``Okx pay singapore: Scan-to-pay with stablecoins at grabpay merchants,'' accessed: Nov. 27, 2025. [Online]. Available: \url{https://www.okx.com/en-au/learn/okx-pay-singapore}
\BIBentrySTDinterwordspacing

\bibitem{businesstimes2025okx}
\BIBentryALTinterwordspacing
{The Business Times (Singapore)}, ``Okx singapore launches stablecoin payments at grabpay merchants,'' accessed: Nov. 27, 2025. [Online]. Available: \url{https://www.businesstimes.com.sg/wealth/crypto-alternative-assets/okx-singapore-launches-stablecoin-payments-grabpay-merchants}
\BIBentrySTDinterwordspacing

\bibitem{cna2025xsgd}
\BIBentryALTinterwordspacing
{Channel NewsAsia}, ``Xsgd launches on coinbase,'' accessed: Oct. 10, 2025. [Online]. Available: \url{https://www.channelnewsasia.com/business/stablecoin-xsgd-singapore-launch-coinbase-crypto-okx-pay-5379656}
\BIBentrySTDinterwordspacing

\bibitem{tian2021enabling}
H.~Tian, K.~Xue, X.~Luo, S.~Li, J.~Xu, J.~Liu, J.~Zhao, and D.~S. Wei, ``Enabling cross-chain transactions: A decentralized cryptocurrency exchange protocol,'' \emph{IEEE Transactions on Information Forensics and Security}, vol.~16, pp. 3928--3941, 2021.

\bibitem{brechlin2024buy}
A.~Brechlin, J.~Sch{\"a}fer, and F.~Armknecht, ``Buy crypto, sell privacy: Investigating the cryptocurrency exchange evonax,'' in \emph{2024 IEEE International Conference on Blockchain and Cryptocurrency}.\hskip 1em plus 0.5em minus 0.4em\relax IEEE, 2024, pp. 1--7.

\bibitem{werner2022sok}
S.~Werner, D.~Perez, L.~Gudgeon, A.~Klages-Mundt, D.~Harz, and W.~Knottenbelt, ``Sok: Decentralized finance (defi),'' in \emph{Proceedings of the 4th ACM Conference on Advances in Financial Technologies}, 2022, pp. 30--46.

\bibitem{zhou2023sok}
L.~Zhou, X.~Xiong, J.~Ernstberger, S.~Chaliasos, Z.~Wang, Y.~Wang, K.~Qin, R.~Wattenhofer, D.~Song, and A.~Gervais, ``Sok: Decentralized finance (defi) attacks,'' in \emph{2023 IEEE Symposium on Security and Privacy}.\hskip 1em plus 0.5em minus 0.4em\relax IEEE, 2023, pp. 2444--2461.

\bibitem{micaReg}
\BIBentryALTinterwordspacing
{European Union}, ``Regulation (eu) 2023/1114 on markets in crypto-assets (mica),'' 2023, accessed Oct. 11, 2025. [Online]. Available: \url{https://eur-lex.europa.eu/eli/reg/2023/1114/oj}
\BIBentrySTDinterwordspacing

\bibitem{US2025GENIUS}
\BIBentryALTinterwordspacing
``Guiding and establishing national innovation for u.s. stablecoins act (genius act),'' 2025, accessed: Nov. 19, 2025. [Online]. Available: \url{https://www.congress.gov/bill/119th-congress/senate-bill/1582/text}
\BIBentrySTDinterwordspacing

\bibitem{neyer2017blockchain}
G.~Neyer and B.~Geva, ``Blockchain and payment systems: What are the benefits and costs?'' \emph{Journal of Payments Strategy \& Systems}, vol.~11, no.~3, pp. 215--225, 2017.

\bibitem{koutrouli2025exploring}
E.~Koutrouli and P.~Manousopoulos, ``Exploring the use of crypto-assets for payments,'' \emph{FinTech}, vol.~4, no.~2, p.~15, 2025.

\bibitem{erdin2021evaluation}
E.~Erdin, S.~Mercan, and K.~Akkaya, ``An evaluation of cryptocurrency payment channel networks and their privacy implications,'' \emph{arXiv preprint arXiv:2102.02659}, 2021.

\bibitem{jonker2019drives}
N.~Jonker, ``What drives the adoption of crypto-payments by online retailers?'' \emph{Electronic Commerce Research and Applications}, vol.~35, p. 100848, 2019.

\bibitem{nabilou2020dark}
H.~Nabilou, ``The dark side of licensing cryptocurrency exchanges as payment institutions,'' \emph{Law and Financial Markets Review}, vol.~14, no.~1, pp. 39--47, 2020.

\bibitem{puschmann2024taxonomy}
T.~Puschmann and M.~Huang-Sui, ``A taxonomy for decentralized finance,'' \emph{International Review of Financial Analysis}, vol.~92, p. 103083, 2024.

\bibitem{worldbank2018gpss}
\BIBentryALTinterwordspacing
{World Bank}, ``Payment systems worldwide: A snapshot—summary outcomes of the fourth global payment systems survey,'' World Bank, Tech. Rep., 2018. [Online]. Available: \url{https://documents1.worldbank.org/curated/en/099011624132054588/pdf/P1787031cca90801019d011a41f21efe348.pdf}
\BIBentrySTDinterwordspacing

\bibitem{ArnerAuerFrost2020}
\BIBentryALTinterwordspacing
D.~Arner, R.~Auer, and J.~Frost, ``Stablecoins: Risks, potential and regulation,'' 2020, accessed: Nov. 19, 2025. [Online]. Available: \url{https://www.bis.org/publ/work905.htm}
\BIBentrySTDinterwordspacing

\bibitem{moin2019sok}
A.~Moin, K.~Sekniqi, and E.~G. Sirer, ``Sok: A classification framework for stablecoin designs,'' in \emph{Financial Cryptography and Data Security}.\hskip 1em plus 0.5em minus 0.4em\relax Springer, 2020, pp. 174--197.

\bibitem{mahrous2025sok}
A.~Mahrous, M.~Caprolu, and R.~Di~Pietro, ``Sok: A structured analysis of economic and technical stablecoin-related research,'' in \emph{2025 IEEE International Conference on Blockchain and Cryptocurrency}.\hskip 1em plus 0.5em minus 0.4em\relax IEEE, 2025, pp. 1--17.

\bibitem{ling2025sok}
S.~Ling, Y.~Du, Y.~Zhou, L.~Wu, C.~Wang, X.~Jia, and H.~Yan, ``Sok: Stablecoin designs, risks, and the stablecoin lego,'' \emph{arXiv preprint arXiv:2506.17622}, 2025.

\bibitem{kahn2009pay}
C.~M. Kahn and W.~Roberds, ``Why pay? an introduction to payments economics,'' \emph{Journal of Financial Intermediation}, vol.~18, no.~1, pp. 1--23, 2009.

\bibitem{ward2006emv}
M.~Ward, ``Emv card payments--an update,'' \emph{Information Security Technical Report}, vol.~11, no.~2, pp. 89--92, 2006.

\bibitem{stearns2011electronic}
D.~L. Stearns, \emph{Electronic value exchange: Origins of the VISA electronic payment system}.\hskip 1em plus 0.5em minus 0.4em\relax Springer, 2011.

\bibitem{ISO_8583_2023}
\BIBentryALTinterwordspacing
{International Organization for Standardization}, ``Iso 8583:2023 — financial-transaction-card-originated messages: Interchange message specifications,'' 2023, accessed: Nov. 27, 2025. [Online]. Available: \url{https://www.iso.org/standard/79451.html}
\BIBentrySTDinterwordspacing

\bibitem{evans2004paying}
D.~S. Evans and R.~Schmalensee, \emph{Paying with plastic: the digital revolution in buying and borrowing}.\hskip 1em plus 0.5em minus 0.4em\relax Mit Press, 2004.

\bibitem{thomadakis2022shortening}
A.~Thomadakis, ``Shortening the settlement cycle: Why europe should not wait too long to introduce t+1,'' \emph{Journal of Securities Operations \& Custody}, vol.~14, no.~4, pp. 350--354, 2022.

\bibitem{benito2025navigating}
J.~Benito, ``Navigating settlement efficiency in a world of csdr and t+1.'' \emph{Journal of Securities Operations \& Custody}, vol.~17, no.~3, 2025.

\bibitem{epc_sctinst}
\BIBentryALTinterwordspacing
{European Payments Council}. {SEPA Instant Credit Transfer}. Accessed: 18 Dec 2025. [Online]. Available: \url{https://www.europeanpaymentscouncil.eu/what-we-do/sepa-instant-credit-transfer}
\BIBentrySTDinterwordspacing

\bibitem{bcb_pix}
\BIBentryALTinterwordspacing
{Banco Central do Brasil}. {Pix}. Accessed: 18 Dec 2025. [Online]. Available: \url{https://www.bcb.gov.br/en/financialstability/pix_en}
\BIBentrySTDinterwordspacing

\bibitem{frb_fednow}
\BIBentryALTinterwordspacing
{Federal Reserve Financial Services}. {FedNow Service}. Accessed: 18 Dec 2025. [Online]. Available: \url{https://www.frbservices.org/financial-services/fednow}
\BIBentrySTDinterwordspacing

\bibitem{BIS2021}
\BIBentryALTinterwordspacing
{Bank for International Settlements}, ``Central bank digital currency: Key features and implications,'' 2021, accessed: Nov. 19, 2025. [Online]. Available: \url{https://www.bis.org/cpmi/publ/d201.pdf}
\BIBentrySTDinterwordspacing

\bibitem{WorldBank2022}
{World Bank Group}, ``Innovations in electronic payment acceptance: Enabling better access for micro and small merchants,'' The World Bank, Washington, DC, Tech. Rep., 2022.

\bibitem{nakamoto2008bitcoin}
S.~Nakamoto and A.~Bitcoin, ``A peer-to-peer electronic cash system,'' \emph{Bitcoin.--URL: https://bitcoin. org/bitcoin. pdf}, vol.~4, no.~2, p.~15, 2008.

\bibitem{Buterin2014}
\BIBentryALTinterwordspacing
V.~Buterin, ``A next-generation smart contract and decentralized application platform,'' 2014, accessed: Oct. 11, 2025. [Online]. Available: \url{https://ethereum.org/en/whitepaper}
\BIBentrySTDinterwordspacing

\bibitem{zheng2020overview}
Z.~Zheng, S.~Xie, H.-N. Dai, W.~Chen, X.~Chen, J.~Weng, and M.~Imran, ``An overview on smart contracts: Challenges, advances and platforms,'' \emph{Future Generation Computer Systems}, vol. 105, pp. 475--491, 2020.

\bibitem{garg2022ethereum}
R.~Garg, ``Ethereum based smart contracts for trade and finance,'' \emph{International Journal of Economics and Management Engineering}, vol.~16, no.~11, pp. 619--629, 2022.

\bibitem{mattke2020cryptocurrency}
J.~Mattke, C.~Maier, and L.~Reis, ``Is cryptocurrency money? three empirical studies analyzing medium of exchange, store of value and unit of account,'' in \emph{Proceedings of the 2020 Computers and People Research Conference}, 2020, pp. 26--35.

\bibitem{TetherTransparency2025}
\BIBentryALTinterwordspacing
Tether, ``Tether transparency: Usdt reserves,'' 2025, accessed: Dec. 4, 2025. [Online]. Available: \url{https://tether.to/en/transparency/?tab=usdt}
\BIBentrySTDinterwordspacing

\bibitem{CircleTransparency2025}
\BIBentryALTinterwordspacing
Circle, ``Circle transparency and reserves reporting,'' 2025, accessed: Dec. 4, 2025. [Online]. Available: \url{https://www.circle.com/transparency}
\BIBentrySTDinterwordspacing

\bibitem{alexandrova2009impact}
B.~Alexandrova-Kabadjova, ``Impact of interchange fees on a nonsaturated multi-agent payment card market,'' \emph{Intelligent Systems in Accounting, Finance and Management: International Journal}, vol.~16, no. 1-2, pp. 33--48, 2009.

\bibitem{cpmi2016glossary}
\BIBentryALTinterwordspacing
{Bank for International Settlements}, ``A glossary of terms used in payments and settlement systems,'' 2016, accessed: Nov. 19, 2025. [Online]. Available: \url{https://www.bis.org/cpmi/publ/d00b.htm}
\BIBentrySTDinterwordspacing

\bibitem{hughes2010developments}
S.~J. Hughes and S.~T. Middlebrook, ``Developments in the laws governing electronic payments made through gift cards, debit and prepaid cards, credit cards, and direct deposits of federal benefits,'' \emph{Bus. Law.}, vol.~66, p. 159, 2010.

\bibitem{humphrey201910}
D.~HUMPHREY, ``10.1 payments overview,'' \emph{The Oxford Handbook of Banking}, p. 285, 2019.

\bibitem{rysman2007empirical}
M.~Rysman, ``An empirical analysis of payment card usage,'' \emph{The Journal of Industrial Economics}, vol.~55, no.~1, pp. 1--36, 2007.

\bibitem{aysan2025local}
A.~F. Aysan, O.~Ozturk, and N.~H. Selim, ``Local power, global reach: The rise and impact of domestic payment networks in the financial landscape,'' \emph{Journal of Science and Technology Policy Management}, vol.~16, no.~1, pp. 143--162, 2025.

\bibitem{ECBGlossary}
\BIBentryALTinterwordspacing
{European Central Bank}. Payments and markets glossary. Accessed: 19 Dec, 2025. [Online]. Available: \url{https://www.ecb.europa.eu/services/glossary/html/act7c.en.html}
\BIBentrySTDinterwordspacing

\bibitem{payuk_fps}
\BIBentryALTinterwordspacing
{Pay.UK}. {Faster Payment System}. Accessed: 18 Dec 2025. [Online]. Available: \url{https://www.wearepay.uk/what-we-do/payment-systems/faster-payment-system/}
\BIBentrySTDinterwordspacing

\bibitem{MoneySmartDirectDebits}
\BIBentryALTinterwordspacing
{Australian Securities and Investments Commission}. Direct debits. MoneySmart (Australian Government). Accessed: 19 Dec, 2025. [Online]. Available: \url{https://moneysmart.gov.au/banking/direct-debits}
\BIBentrySTDinterwordspacing

\bibitem{epc_sdd}
\BIBentryALTinterwordspacing
{European Payments Council}. {SEPA Direct Debit}. Accessed: 18 Dec 2025. [Online]. Available: \url{https://www.europeanpaymentscouncil.eu/what-we-do/sepa-direct-debit}
\BIBentrySTDinterwordspacing

\bibitem{bacs_dd}
\BIBentryALTinterwordspacing
{Bacs}. {Direct Debit}. Accessed: 18 Dec 2025. [Online]. Available: \url{https://www.bacs.co.uk/bacs-schemes/direct-debit}
\BIBentrySTDinterwordspacing

\bibitem{lilge2001evolution}
M.~Lilge, ``Evolution of prepaid service towards a real-time payment system,'' in \emph{IEEE Intelligent Network 2001 Workshop. IN 2001 Conference Record (Cat. No. 01TH8566)}.\hskip 1em plus 0.5em minus 0.4em\relax IEEE, 2001, pp. 195--198.

\bibitem{horne2016gift}
D.~Horne and N.~Bendle, ``Gift cards: a review and research agenda,'' \emph{The international review of retail, distribution and consumer research}, vol.~26, no.~2, pp. 154--170, 2016.

\bibitem{prager2009interchange}
R.~A. Prager, M.~D. Manuszak, E.~K. Kiser, and R.~Borzekowski, ``Interchange fees and payment card networks: Economics, industry developments, and policy issues,'' 2009.

\bibitem{chakravorti2003theory}
S.~Chakravorti, ``Theory of credit card networks: A survey of the literature,'' \emph{Review of network Economics}, vol.~2, no.~2, 2003.

\bibitem{scardovi2017digital}
C.~Scardovi, ``Digital transformation in payments,'' in \emph{Digital Transformation in Financial Services}.\hskip 1em plus 0.5em minus 0.4em\relax Springer, 2017, pp. 65--84.

\bibitem{giardina1993merchant}
A.~Giardina, ``The merchant,'' \emph{The Romans}, pp. 245--271, 1993.

\bibitem{worldpay}
\BIBentryALTinterwordspacing
{Worldpay}. {Worldpay}. Accessed: 18 Dec 2025. [Online]. Available: \url{https://www.worldpay.com/en-GB}
\BIBentrySTDinterwordspacing

\bibitem{stripe_au}
\BIBentryALTinterwordspacing
{Stripe}. {Stripe Australia}. Accessed: 18 Dec 2025. [Online]. Available: \url{https://stripe.com/au}
\BIBentrySTDinterwordspacing

\bibitem{adyen_au}
\BIBentryALTinterwordspacing
{Adyen}. {Adyen Australia}. Accessed: 18 Dec 2025. [Online]. Available: \url{https://www.adyen.com/en_AU}
\BIBentrySTDinterwordspacing

\bibitem{guo2012interchange}
H.~Guo, M.~Leng, and Y.~Wang, ``Interchange fee rate, merchant discount rate, and retail price in a credit card network: A game-theoretic analysis,'' \emph{Naval Research Logistics (NRL)}, vol.~59, no.~7, pp. 525--551, 2012.

\bibitem{unionpay_intl}
\BIBentryALTinterwordspacing
{UnionPay International}. {UnionPay International}. Accessed: 18 Dec 2025. [Online]. Available: \url{https://www.unionpayintl.com/en/}
\BIBentrySTDinterwordspacing

\bibitem{rba_scheme_fees}
\BIBentryALTinterwordspacing
{Reserve Bank of Australia}. {Backgrounder on Interchange and Scheme Fees}. Accessed: 18 Dec 2025. [Online]. Available: \url{https://www.rba.gov.au/payments-and-infrastructure/review-of-retail-payments-regulation/backgrounders/backgrounder-on-interchange-and-scheme-fees.html}
\BIBentrySTDinterwordspacing

\bibitem{verifone_poi}
\BIBentryALTinterwordspacing
{Verifone}. {Points of Interaction | Verifone Central}. Accessed: 18 Dec 2025. [Online]. Available: \url{https://verifone.cloud/docs/verifone-central/verifone-central/manage-your-account/administration/points-interaction}
\BIBentrySTDinterwordspacing

\bibitem{stripe2023}
\BIBentryALTinterwordspacing
I.~Stripe, ``How payment transaction processing works,'' 2023, accessed: 21 Dec, 2025. [Online]. Available: \url{https://stripe.com/au/resources/more/how-payment-transaction-processing-works}
\BIBentrySTDinterwordspacing

\bibitem{visaCoreRules2025}
\BIBentryALTinterwordspacing
{Visa Inc.}, ``Visa core rules and visa product and service rules,'' 2025, effective May 20, 2025; Accessed Oct. 11, 2025. [Online]. Available: \url{https://usa.visa.com/dam/VCOM/download/about-visa/visa-rules-public.pdf}
\BIBentrySTDinterwordspacing

\bibitem{checkout2024_cardAuthorization}
\BIBentryALTinterwordspacing
{Checkout.com}, ``Card authorization explained: How does it work?'' 2024, accessed: Nov. 27, 2025. [Online]. Available: \url{https://www.checkout.com/blog/card-authorization-explained}
\BIBentrySTDinterwordspacing

\bibitem{thrivent2025}
\BIBentryALTinterwordspacing
Thrivent, ``T+1 settlement: Definitions, pros, cons \& why it's important for investors,'' 2025, accessed: 21 Dec, 2025. [Online]. Available: \url{https://www.thrivent.com/insights/investing/t-1-settlement-definitions-pros-cons-why-its-important-for-investors}
\BIBentrySTDinterwordspacing

\bibitem{kahn2002economics}
C.~M. Kahn and W.~Roberds, ``The economics of payment finality,'' \emph{Economic Review-Federal Reserve Bank of Atlanta}, vol.~87, no.~2, pp. 1--12, 2002.

\bibitem{RBA2022}
\BIBentryALTinterwordspacing
C.~Dark, E.~Rogerson, N.~Rowbotham, and P.~Wallis, ``Stablecoins: Market developments, risks and regulation,'' \emph{Bulletin}, Dec 2022. [Online]. Available: \url{https://www.rba.gov.au/publications/bulletin/2022/dec/pdf/stablecoins-market-developments-risks-and-regulation.pdf}
\BIBentrySTDinterwordspacing

\bibitem{ferdous2021survey}
M.~S. Ferdous, M.~J.~M. Chowdhury, and M.~A. Hoque, ``A survey of consensus algorithms in public blockchain systems for crypto-currencies,'' \emph{Journal of Network and Computer Applications}, vol. 182, p. 103035, 2021.

\bibitem{yakovenko2018solana}
A.~Yakovenko, ``Solana: A new architecture for a high performance blockchain v0. 8.13,'' 2018.

\bibitem{asayag2018fair}
A.~Asayag, G.~Cohen, I.~Grayevsky, M.~Leshkowitz, O.~Rottenstreich, R.~Tamari, and D.~Yakira, ``A fair consensus protocol for transaction ordering,'' in \emph{2018 IEEE 26th International Conference on Network Protocols}.\hskip 1em plus 0.5em minus 0.4em\relax IEEE, 2018, pp. 55--65.

\bibitem{das2019transaction}
R.~A. Das, M.~M.~S. Pahalovi, and M.~N. Yanhaona, ``Transaction finality through ledger checkpoints,'' in \emph{2019 IEEE 25th International Conference on Parallel and Distributed Systems}.\hskip 1em plus 0.5em minus 0.4em\relax IEEE, 2019, pp. 183--192.

\bibitem{malamud2017decentralized}
S.~Malamud and M.~Rostek, ``Decentralized exchange,'' \emph{American Economic Review}, vol. 107, no.~11, pp. 3320--3362, 2017.

\bibitem{kaplan2023decentralized}
B.~Kaplan, V.~F. Benli, and E.~A. Alp, ``Decentralized finance and new lending protocols,'' \emph{PressAcademia Procedia}, vol.~16, no.~1, pp. 192--195, 2023.

\bibitem{li2025blockchain}
N.~Li, M.~Qi, Z.~Xu, X.~Zhu, W.~Zhou, S.~Wen, and Y.~Xiang, ``Blockchain cross-chain bridge security: Challenges, solutions, and future outlook,'' \emph{Distributed Ledger Technologies: Research and Practice}, vol.~4, no.~1, pp. 1--34, 2025.

\bibitem{suratkar2020cryptocurrency}
S.~Suratkar, M.~Shirole, and S.~Bhirud, ``Cryptocurrency wallet: A review,'' in \emph{2020 4th international conference on computer, communication and signal processing}.\hskip 1em plus 0.5em minus 0.4em\relax IEEE, 2020, pp. 1--7.

\bibitem{dotan2021survey}
M.~Dotan, Y.-A. Pignolet, S.~Schmid, S.~Tochner, and A.~Zohar, ``Survey on blockchain networking: Context, state-of-the-art, challenges,'' \emph{ACM Computing Surveys}, vol.~54, no.~5, pp. 1--34, 2021.

\bibitem{kazerani2017determining}
A.~Kazerani, D.~Rosati, and B.~Lesser, ``Determining the usability of bitcoin for beginners using change tip and coinbase,'' in \emph{Proceedings of the 35th ACM International Conference on the Design of Communication}, 2017, pp. 1--5.

\bibitem{trautman2015commerce}
L.~J. Trautman, ``E-commerce, cyber, and electronic payment system risks: lessons from paypal,'' \emph{UC Davis Bus. LJ}, vol.~16, p. 261, 2015.

\bibitem{chen2020you}
T.-H. Chen, ``Do you know your customer? bank risk assessment based on machine learning,'' \emph{Applied Soft Computing}, vol.~86, p. 105779, 2020.

\bibitem{lee2023using}
W.-M. Lee, ``Using the metamask crypto-wallet,'' in \emph{Beginning Ethereum Smart Contracts Programming: With Examples in Python, Solidity, and JavaScript}.\hskip 1em plus 0.5em minus 0.4em\relax Springer, 2023, pp. 111--144.

\bibitem{PhantomWallet2025}
\BIBentryALTinterwordspacing
Phantom. (2025) Phantom — the crypto wallet for everyone. Accessed: Dec. 4, 2025. [Online]. Available: \url{https://phantom.com}
\BIBentrySTDinterwordspacing

\bibitem{Circle2025}
\BIBentryALTinterwordspacing
{Circle Internet Group, Inc.}, ``Circle — open infrastructure for faster, smarter payments,'' 2025, accessed: 20 Dec, 2025. [Online]. Available: \url{https://www.circle.com}
\BIBentrySTDinterwordspacing

\bibitem{Tether2025}
\BIBentryALTinterwordspacing
{Tether Limited}, ``Tether — official home of tether,'' 2025, accessed: 20 Dec, 2025. [Online]. Available: \url{https://tether.to/en}
\BIBentrySTDinterwordspacing

\bibitem{qin2021cefi}
K.~Qin, L.~Zhou, Y.~Afonin, L.~Lazzaretti, and A.~Gervais, ``Cefi vs. defi--comparing centralized to decentralized finance,'' \emph{arXiv preprint arXiv:2106.08157}, 2021.

\bibitem{arslanian2022stablecoins}
H.~Arslanian, ``Stablecoins,'' in \emph{The Book of Crypto: The Complete Guide to Understanding Bitcoin, Cryptocurrencies and Digital Assets}.\hskip 1em plus 0.5em minus 0.4em\relax Springer, 2022, pp. 149--170.

\bibitem{xu2023survey}
J.~Xu, C.~Wang, and X.~Jia, ``A survey of blockchain consensus protocols,'' \emph{ACM Computing Surveys}, vol.~55, no. 13s, pp. 1--35, 2023.

\bibitem{nabilou2024probabilistic}
H.~Nabilou, ``Probabilistic settlement finality in proof-of-work blockchains: Legal considerations,'' \emph{Bus. \& Fin. L. Rev.}, vol.~7, p. 139, 2024.

\bibitem{gramoli2020blockchain}
V.~Gramoli, ``From blockchain consensus back to byzantine consensus,'' \emph{Future Generation Computer Systems}, vol. 107, pp. 760--769, 2020.

\bibitem{xiao2020survey}
Y.~Xiao, N.~Zhang, W.~Lou, and Y.~T. Hou, ``A survey of distributed consensus protocols for blockchain networks,'' \emph{IEEE communications surveys \& tutorials}, vol.~22, no.~2, pp. 1432--1465, 2020.

\bibitem{politou2019blockchain}
E.~Politou, F.~Casino, E.~Alepis, and C.~Patsakis, ``Blockchain mutability: Challenges and proposed solutions,'' \emph{IEEE Transactions on Emerging Topics in Computing}, vol.~9, no.~4, pp. 1972--1986, 2019.

\bibitem{ye2023survey}
T.~Ye, M.~Luo, Y.~Yang, K.-K.~R. Choo, and D.~He, ``A survey on redactable blockchain: Challenges and opportunities,'' \emph{IEEE Transactions on Network Science and Engineering}, vol.~10, no.~3, pp. 1669--1683, 2023.

\bibitem{johnson2020decentralized}
K.~N. Johnson, ``Decentralized finance: Regulating cryptocurrency exchanges,'' \emph{Wm. \& Mary L. Rev.}, vol.~62, p. 1911, 2020.

\bibitem{okasova2024using}
K.~Okasov{\'a} and K.~Ko{\v{s}}t'{\'a}l, ``Using machine learning for predicting arbitrage occurrences in cryptocurrency exchanges,'' in \emph{2024 IEEE International Conference on Blockchain and Cryptocurrency}.\hskip 1em plus 0.5em minus 0.4em\relax IEEE, 2024, pp. 1--7.

\bibitem{CircleUSDC2025}
\BIBentryALTinterwordspacing
{Circle Internet Group, Inc.}, ``Usdc redemption structure,'' 2025, accessed: 20 Dec, 2025. [Online]. Available: \url{https://help.circle.com/s/article/USDC-redemption-structure?language=en_US&category=Fees_and_Billing}
\BIBentrySTDinterwordspacing

\bibitem{TetherRedeem2025}
\BIBentryALTinterwordspacing
{Tether Operations, S.A. de C.V.}, ``Redeem tether tokens to fiat currency,'' 2025, accessed: 20 Dec, 2025. [Online]. Available: \url{https://tether.to/ru/redeem-tethers-to-fiat-currency}
\BIBentrySTDinterwordspacing

\bibitem{houy2023security}
S.~Houy, P.~Schmid, and A.~Bartel, ``Security aspects of cryptocurrency wallets—a systematic literature review,'' \emph{ACM Computing Surveys}, vol.~56, no.~1, pp. 1--31, 2023.

\bibitem{puthal2018everything}
D.~Puthal, N.~Malik, S.~P. Mohanty, E.~Kougianos, and G.~Das, ``Everything you wanted to know about the blockchain: Its promise, components, processes, and problems,'' \emph{IEEE Consumer Electronics Magazine}, vol.~7, no.~4, pp. 6--14, 2018.

\bibitem{lashkari2021comprehensive}
B.~Lashkari and P.~Musilek, ``A comprehensive review of blockchain consensus mechanisms,'' \emph{IEEE access}, vol.~9, pp. 43\,620--43\,652, 2021.

\bibitem{ruan2020transactional}
P.~Ruan, D.~Loghin, Q.-T. Ta, M.~Zhang, G.~Chen, and B.~C. Ooi, ``A transactional perspective on execute-order-validate blockchains,'' in \emph{Proceedings of the 2020 ACM SIGMOD International Conference on Management of Data}, 2020, pp. 543--557.

\bibitem{salau2022making}
A.~Salau, R.~Dantu, K.~Morozov, S.~Badruddoja, and K.~Upadhyay, ``Making blockchain validators honest,'' in \emph{2022 Fourth International Conference on Blockchain Computing and Applications (BCCA)}.\hskip 1em plus 0.5em minus 0.4em\relax IEEE, 2022, pp. 267--273.

\bibitem{zhang2020overview}
C.~Zhang, C.~Wu, and X.~Wang, ``Overview of blockchain consensus mechanism,'' in \emph{Proceedings of the 2020 2nd International Conference on Big Data Engineering}, 2020, pp. 7--12.

\bibitem{zhang2021transaction}
L.~Zhang, R.~Zhou, Q.~Liu, J.~Xu, and C.~Liu, ``Transaction confirmation time estimation in the bitcoin blockchain,'' in \emph{International Conference on Web Information Systems Engineering}.\hskip 1em plus 0.5em minus 0.4em\relax Springer, 2021, pp. 30--45.

\bibitem{asante2021distributed}
M.~Asante, G.~Epiphaniou, C.~Maple, H.~Al-Khateeb, M.~Bottarelli, and K.~Z. Ghafoor, ``Distributed ledger technologies in supply chain security management: A comprehensive survey,'' \emph{IEEE Transactions on Engineering Management}, vol.~70, no.~2, pp. 713--739, 2021.

\bibitem{liu2021characterizing}
L.~Liu, L.~Wei, W.~Zhang, M.~Wen, Y.~Liu, and S.-C. Cheung, ``Characterizing transaction-reverting statements in ethereum smart contracts,'' in \emph{2021 36th IEEE/ACM International Conference on Automated Software Engineering (ASE)}.\hskip 1em plus 0.5em minus 0.4em\relax IEEE, 2021, pp. 630--641.

\bibitem{liao2023smartstate}
Z.~Liao, S.~Hao, Y.~Nan, and Z.~Zheng, ``Smartstate: Detecting state-reverting vulnerabilities in smart contracts via fine-grained state-dependency analysis,'' in \emph{Proceedings of the 32nd ACM SIGSOFT International Symposium on Software Testing and Analysis}, 2023, pp. 980--991.

\bibitem{rysman2009economics}
M.~Rysman, ``The economics of two-sided markets,'' \emph{Journal of economic perspectives}, vol.~23, no.~3, pp. 125--143, 2009.

\bibitem{rochet2002cooperation}
J.-C. Rochet and J.~Tirole, ``Cooperation among competitors: Some economics of payment card associations,'' \emph{Rand Journal of economics}, pp. 549--570, 2002.

\bibitem{helcim2024}
\BIBentryALTinterwordspacing
Helcim, ``Merchant discount rate: How to calculate,'' 2024, accessed: 23 Dec, 2025. [Online]. Available: \url{https://www.helcim.com/guides/merchant-discount-rate/}
\BIBentrySTDinterwordspacing

\bibitem{investopedia2025}
\BIBentryALTinterwordspacing
Investopedia, ``Understanding merchant discount rate: Definition \& key fees explained,'' 2025, accessed: 23 Dec, 2025. [Online]. Available: \url{https://www.investopedia.com/terms/m/merchant-discount-rate.asp}
\BIBentrySTDinterwordspacing

\bibitem{EUReg2015}
\BIBentryALTinterwordspacing
E.~Commission, ``Regulation (eu) 2015/751 of the european parliament and of the council of 29 april 2015 on interchange fees for card-based payment transactions,'' 2015, accessed: Nov. 19, 2025. [Online]. Available: \url{https://eur-lex.europa.eu/eli/reg/2015/751/oj/eng}
\BIBentrySTDinterwordspacing

\bibitem{FederalReserve2024}
\BIBentryALTinterwordspacing
F.~Reserve, ``Regulation ii - average debit card interchange fee by payment card network,'' 2024, accessed: Nov. 19, 2025. [Online]. Available: \url{https://www.federalreserve.gov/paymentsystems/regii-average-interchange-fee.htm}
\BIBentrySTDinterwordspacing

\bibitem{BankCanada2022}
\BIBentryALTinterwordspacing
A.~Ho, S.~Darbha, Y.~Gorelkina, and A.~García, ``The relative benefits and risks of stablecoins as a means of payment: A case study perspective,'' December 2022, accessed: Nov. 19, 2025. [Online]. Available: \url{https://www.bankofcanada.ca/2022/12/staff-discussion-paper-2022-21}
\BIBentrySTDinterwordspacing

\bibitem{donmez2022transaction}
A.~Donmez and A.~Karaivanov, ``Transaction fee economics in the ethereum blockchain,'' \emph{Economic Inquiry}, vol.~60, no.~1, pp. 265--292, 2022.

\bibitem{gudgeon2020sok}
L.~Gudgeon, P.~Moreno-Sanchez, S.~Roos, P.~McCorry, and A.~Gervais, ``Sok: Layer-two blockchain protocols,'' in \emph{International Conference on Financial Cryptography and Data Security}.\hskip 1em plus 0.5em minus 0.4em\relax Springer, 2020, pp. 201--226.

\bibitem{sguanci2021layer}
C.~Sguanci, R.~Spatafora, and A.~M. Vergani, ``Layer 2 blockchain scaling: A survey,'' \emph{arXiv preprint arXiv:2107.10881}, 2021.

\bibitem{CoinLaw2025L2Fees}
\BIBentryALTinterwordspacing
CoinLaw.io, ``Gas fee markets on layer-2 statistics 2025: Key insights,'' 2025. [Online]. Available: \url{https://coinlaw.io/gas-fee-markets-on-layer-2-statistics}
\BIBentrySTDinterwordspacing

\bibitem{Li2024MStableChain}
M.~Li, B.~Gao, K.~Toyoda, Y.~Yang, J.~Samsudin, H.~Zhang, S.~Lu, T.~H. Tng, K.~Choo, and Q.~Wei, ``Mstablechain: Towards multi-native stablecoins in evm-compatible blockchain for stable fee and mass adoption,'' \emph{arXiv preprint arXiv:2410.22100}, 2024.

\bibitem{EuropeanCommission2020IFR}
{European Commission}, ``Report on the application of regulation (eu) 2015/751 on interchange fees for card-based payment transactions,'' European Commission, Commission Staff Working Document SWD(2020) 118 final, June 2020.

\bibitem{VisaISA2023}
\BIBentryALTinterwordspacing
{Tidal Commerce}, ``Isa fee explained: Visa international service assessment,'' 2023, accessed: Dec. 5, 2025. [Online]. Available: \url{https://www.tidalcommerce.com/learn/isa-fee}
\BIBentrySTDinterwordspacing

\bibitem{StripePricing2025}
\BIBentryALTinterwordspacing
{Stripe}, ``Pricing and fees: International card costs,'' 2025, accessed: Dec. 5, 2025. [Online]. Available: \url{https://stripe.com/pricing}
\BIBentrySTDinterwordspacing

\bibitem{circleNextChapter}
\BIBentryALTinterwordspacing
{Circle Internet Financial}, ``Ushering in the next chapter for usdc,'' 2023, centre governance brought in-house; Accessed Oct. 11, 2025. [Online]. Available: \url{https://www.circle.com/blog/ushering-in-the-next-chapter-for-usdc}
\BIBentrySTDinterwordspacing

\bibitem{ferrin2002total}
B.~G. Ferrin and R.~E. Plank, ``Total cost of ownership models: An exploratory study,'' \emph{Journal of Supply chain management}, vol.~38, no.~2, pp. 18--29, 2002.

\bibitem{zetzsche2021markets}
D.~A. Zetzsche, F.~Annunziata, D.~W. Arner, and R.~P. Buckley, ``The markets in crypto-assets regulation (mica) and the eu digital finance strategy,'' \emph{Capital Markets Law Journal}, vol.~16, no.~2, pp. 203--225, 2021.

\bibitem{mas2023stablecoin}
\BIBentryALTinterwordspacing
{Monetary Authority of Singapore}, ``Mas finalises stablecoin regulatory framework,'' August 2023, mandates 100\% liquid asset backing and redemption within 5 business days. [Online]. Available: \url{https://www.mas.gov.sg/news/media-releases/2023/mas-finalises-stablecoin-regulatory-framework}
\BIBentrySTDinterwordspacing

\bibitem{hkma2024sandbox}
{Hong Kong Monetary Authority}, ``Conclusion of discussion paper on crypto-assets and stablecoins,'' HKMA, Tech. Rep., 2023, establishes the mandatory licensing regime for FRS issuers.

\bibitem{EUDirective}
\BIBentryALTinterwordspacing
{European Parliament and Council of the European Union}, ``Directive 98/26/ec of the european parliament and of the council of 19 may 1998 on settlement finality in payment and securities settlement systems,'' 2024, accessed: Nov. 19, 2025. [Online]. Available: \url{https://eur-lex.europa.eu/eli/dir/1998/26/oj/eng}
\BIBentrySTDinterwordspacing

\bibitem{gervais2016security}
A.~Gervais, G.~O. Karame, K.~W{\"u}st, V.~Glykantzis, H.~Ritzdorf, and S.~Capkun, ``On the security and performance of proof of work blockchains,'' in \emph{Proceedings of the 2016 ACM SIGSAC conference on computer and communications security}, 2016, pp. 3--16.

\bibitem{nguyen2019proof}
C.~T. Nguyen, D.~T. Hoang, D.~N. Nguyen, D.~Niyato, H.~T. Nguyen, and E.~Dutkiewicz, ``Proof-of-stake consensus mechanisms for future blockchain networks: fundamentals, applications and opportunities,'' \emph{IEEE access}, vol.~7, pp. 85\,727--85\,745, 2019.

\bibitem{GDPRArticle17}
\BIBentryALTinterwordspacing
``General data protection regulation, article 17: Right to erasure,'' 2016. [Online]. Available: \url{https://gdpr-info.eu/art-17-gdpr}
\BIBentrySTDinterwordspacing

\bibitem{liang2024identity}
W.~Liang, Y.~Liu, C.~Yang, S.~Xie, K.~Li, and W.~Susilo, ``On identity, transaction, and smart contract privacy on permissioned and permissionless blockchain: a comprehensive survey,'' \emph{ACM Computing Surveys}, vol.~56, no.~12, pp. 1--35, 2024.

\bibitem{biryukov2019deanonymization}
A.~Biryukov and S.~Tikhomirov, ``Deanonymization and linkability of cryptocurrency transactions based on network analysis,'' in \emph{2019 IEEE European symposium on security and privacy}.\hskip 1em plus 0.5em minus 0.4em\relax IEEE, 2019, pp. 172--184.

\bibitem{xiang2023babd}
Y.~Xiang, Y.~Lei, D.~Bao, T.~Li, Q.~Yang, W.~Liu, W.~Ren, and K.-K.~R. Choo, ``Babd: A bitcoin address behavior dataset for pattern analysis,'' \emph{IEEE Transactions on Information Forensics and Security}, vol.~19, pp. 2171--2185, 2023.

\bibitem{finck2018blockchains}
M.~Finck, ``Blockchains and data protection in the european union,'' \emph{Eur. Data Prot. L. Rev.}, vol.~4, p.~17, 2018.

\bibitem{fiege1987zero}
U.~Fiege, A.~Fiat, and A.~Shamir, ``Zero knowledge proofs of identity,'' in \emph{Proceedings of the nineteenth annual ACM symposium on Theory of computing}, 1987, pp. 210--217.

\bibitem{sun2021survey}
X.~Sun, F.~R. Yu, P.~Zhang, Z.~Sun, W.~Xie, and X.~Peng, ``A survey on zero-knowledge proof in blockchain,'' \emph{IEEE Network}, vol.~35, no.~4, pp. 198--205, 2021.

\bibitem{lindell2020secure}
Y.~Lindell, ``Secure multiparty computation,'' \emph{Communications of the ACM}, vol.~64, no.~1, pp. 86--96, 2020.

\bibitem{zhong2019secure}
H.~Zhong, Y.~Sang, Y.~Zhang, and Z.~Xi, ``Secure multi-party computation on blockchain: An overview,'' in \emph{International symposium on parallel architectures, algorithms and programming}.\hskip 1em plus 0.5em minus 0.4em\relax Springer, 2019, pp. 452--460.

\bibitem{jauernig2020trusted}
P.~Jauernig, A.-R. Sadeghi, and E.~Stapf, ``Trusted execution environments: properties, applications, and challenges,'' \emph{IEEE Security \& Privacy}, vol.~18, no.~2, pp. 56--60, 2020.

\bibitem{lind2017teechain}
J.~Lind, I.~Eyal, F.~Kelbert, O.~Naor, P.~Pietzuch, and E.~G. Sirer, ``Teechain: Scalable blockchain payments using trusted execution environments,'' \emph{arXiv preprint arXiv:1707.05454}, 2017.

\bibitem{marangunic2015technology}
N.~Maranguni{\'c} and A.~Grani{\'c}, ``Technology acceptance model: a literature review from 1986 to 2013,'' \emph{Universal access in the information society}, vol.~14, no.~1, pp. 81--95, 2015.

\bibitem{venkatesh2016unified}
V.~Venkatesh and J.~Thong, ``Unified theory of acceptance and use of technology: A synthesis and the road ahead,'' \emph{Journal of the association for Information Systems}, 2016.

\bibitem{quan2023mobile}
W.~Quan, H.~Moon, S.~S. Kim, and H.~Han, ``Mobile, traditional, and cryptocurrency payments influence consumer trust, attitude, and destination choice: Chinese versus koreans,'' \emph{International Journal of Hospitality Management}, vol. 108, p. 103363, 2023.

\bibitem{sangari2024driving}
M.~S. Sangari and A.~Mashatan, ``What is driving consumer resistance to crypto-payment? a multianalytical investigation,'' \emph{Psychology \& Marketing}, vol.~41, no.~3, pp. 575--591, 2024.

\bibitem{si2024understanding}
J.~J. Si, T.~Sharma, and K.~Y. Wang, ``Understanding user-perceived security risks and mitigation strategies in the web3 ecosystem,'' in \emph{Proceedings of the 2024 CHI Conference on Human Factors in Computing Systems}, 2024, pp. 1--22.

\bibitem{guan2025security}
M.~Y. Guan, Y.~Yu, T.~Sharma, M.~Z. Huang, K.~Qin, Y.~Wang, and K.~Y. Wang, ``Security perceptions of users in stablecoins: Advantages and risks within the cryptocurrency ecosystem,'' in \emph{2025 IEEE Symposium on Security and Privacy (SP)}.\hskip 1em plus 0.5em minus 0.4em\relax IEEE, 2025, pp. 2753--2771.

\bibitem{henry2018bitcoin}
C.~S. Henry, K.~P. Huynh, and G.~Nicholls, ``Bitcoin awareness and usage in canada,'' \emph{Journal of Digital Banking}, vol.~2, no.~4, pp. 311--337, 2018.

\bibitem{schuh2016us}
S.~Schuh and O.~Shy, ``Us consumers’ adoption and use of bitcoin and other virtual currencies,'' in \emph{DeNederlandsche bank, Conference entitled “Retail payments: mapping out the road ahead}, 2016.

\bibitem{shahzad2018empirical}
F.~Shahzad, G.~Xiu, J.~Wang, and M.~Shahbaz, ``An empirical investigation on the adoption of cryptocurrencies among the people of mainland china,'' \emph{Technology in Society}, vol.~55, pp. 33--40, 2018.

\bibitem{polasik2015price}
M.~Polasik, A.~I. Piotrowska, T.~P. Wisniewski, R.~Kotkowski, and G.~Lightfoot, ``Price fluctuations and the use of bitcoin: An empirical inquiry,'' \emph{International journal of electronic commerce}, vol.~20, no.~1, pp. 9--49, 2015.

\bibitem{silinskyte2014understanding}
J.~Silinskyte, ``Understanding bitcoin adoption: Unified theory of acceptance and use of technology (utaut) application (master’s thesis),'' \emph{University Leiden}, 2014.

\bibitem{malakhov2023analysis}
I.~Malakhov, A.~Marin, and S.~Rossi, ``Analysis of the confirmation time in proof-of-work blockchains,'' \emph{Future Generation Computer Systems}, vol. 147, pp. 275--291, 2023.

\bibitem{kawase2020priority}
Y.~Kawase and S.~Kasahara, ``Priority queueing analysis of transaction-confirmation time for bitcoin.'' \emph{Journal of Industrial \& Management Optimization}, vol.~16, no.~3, 2020.

\bibitem{moore2010economics}
T.~Moore, ``The economics of cybersecurity: Principles and policy options,'' \emph{International Journal of Critical Infrastructure Protection}, vol.~3, no. 3-4, pp. 103--117, 2010.

\bibitem{gramoli2024stabl}
V.~Gramoli, R.~Guerraoui, A.~Lebedev, and G.~Voron, ``Stabl: Blockchain fault tolerance,'' \emph{arXiv preprint arXiv:2409.13142}, 2024.

\bibitem{castro1999practical}
M.~Castro, B.~Liskov \emph{et~al.}, ``Practical byzantine fault tolerance,'' in \emph{USENIX Symposium on Operating Systems Design and Implementation}, vol.~99, no. 1999, 1999, pp. 173--186.

\bibitem{sousa2018byzantine}
J.~Sousa, A.~Bessani, and M.~Vukolic, ``A byzantine fault-tolerant ordering service for the hyperledger fabric blockchain platform,'' in \emph{2018 48th annual IEEE/IFIP international conference on dependable systems and networks}.\hskip 1em plus 0.5em minus 0.4em\relax IEEE, 2018, pp. 51--58.

\bibitem{bohme2015bitcoin}
R.~B{\"o}hme, N.~Christin, B.~Edelman, and T.~Moore, ``Bitcoin: Economics, technology, and governance,'' \emph{Journal of economic Perspectives}, vol.~29, no.~2, pp. 213--238, 2015.

\bibitem{motepalli2023sok}
S.~Motepalli, L.~Freitas, and B.~Livshits, ``Sok: Decentralized sequencers for rollups,'' \emph{arXiv preprint arXiv:2310.03616}, 2023.

\bibitem{thibault2022blockchain}
L.~T. Thibault, T.~Sarry, and A.~S. Hafid, ``Blockchain scaling using rollups: A comprehensive survey,'' \emph{IEEE Access}, vol.~10, pp. 93\,039--93\,054, 2022.

\bibitem{mccorry2021sok}
P.~McCorry, C.~Buckland, B.~Yee, and D.~Song, ``Sok: Validating bridges as a scaling solution for blockchains,'' \emph{Cryptology ePrint Archive}, 2021.

\bibitem{anderson2010security}
R.~Anderson, \emph{Security engineering: a guide to building dependable distributed systems}.\hskip 1em plus 0.5em minus 0.4em\relax John Wiley \& Sons, 2010.

\bibitem{morse2008pci}
E.~A. Morse and V.~Raval, ``Pci dss: Payment card industry data security standards in context,'' \emph{Computer Law \& Security Review}, vol.~24, no.~6, pp. 540--554, 2008.

\bibitem{karantias2020sok}
K.~Karantias, ``Sok: A taxonomy of cryptocurrency wallets,'' \emph{Cryptology ePrint Archive}, 2020.

\bibitem{he2020smart}
D.~He, Z.~Deng, Y.~Zhang, S.~Chan, Y.~Cheng, and N.~Guizani, ``Smart contract vulnerability analysis and security audit,'' \emph{IEEE Network}, vol.~34, no.~5, pp. 276--282, 2020.

\bibitem{zhuang2021smart}
Y.~Zhuang, Z.~Liu, P.~Qian, Q.~Liu, X.~Wang, and Q.~He, ``Smart contract vulnerability detection using graph neural networks,'' in \emph{Proceedings of the twenty-ninth international conference on international joint conferences on artificial intelligence}, 2021, pp. 3283--3290.

\bibitem{chen2025chatgpt}
C.~Chen, J.~Su, J.~Chen, Y.~Wang, T.~Bi, J.~Yu, Y.~Wang, X.~Lin, T.~Chen, and Z.~Zheng, ``When chatgpt meets smart contract vulnerability detection: How far are we?'' \emph{ACM Transactions on Software Engineering and Methodology}, vol.~34, no.~4, pp. 1--30, 2025.

\bibitem{katz1985network}
M.~L. Katz and C.~Shapiro, ``Network externalities, competition, and compatibility,'' \emph{The American economic review}, vol.~75, no.~3, pp. 424--440, 1985.

\bibitem{farrell1986installed}
J.~Farrell and G.~Saloner, ``Installed base and compatibility: Innovation, product preannouncements, and predation,'' \emph{The American economic review}, pp. 940--955, 1986.

\bibitem{nilsonreport2024}
\BIBentryALTinterwordspacing
{Nilson}, ``Worldwide card acceptance locations for global brands midyear 2024,'' 2024, accessed: Nov. 27, 2025. [Online]. Available: \url{https://nilsonreport.com/articles/worldwide-card-acceptance-locations-for-global-brands-midyear-2024}
\BIBentrySTDinterwordspacing

\bibitem{li2025blockchain1}
C.~Li, R.~Xu, B.~Palanisamy, L.~Duan, M.~Shen, J.~Liu, and W.~Wang, ``Blockchain takeovers in web 3.0: An empirical study on the tron-steem incident,'' \emph{ACM Transactions on the Web}, vol.~19, no.~2, pp. 1--23, 2025.

\bibitem{xie2022zkbridge}
T.~Xie, J.~Zhang, Z.~Cheng, F.~Zhang, Y.~Zhang, Y.~Jia, D.~Boneh, and D.~Song, ``zkbridge: Trustless cross-chain bridges made practical,'' in \emph{Proceedings of the 2022 ACM SIGSAC Conference on Computer and Communications Security}, 2022, pp. 3003--3017.

\bibitem{lee2023sok}
S.-S. Lee, A.~Murashkin, M.~Derka, and J.~Gorzny, ``Sok: Not quite water under the bridge: Review of cross-chain bridge hacks,'' in \emph{2023 IEEE International Conference on Blockchain and Cryptocurrency}.\hskip 1em plus 0.5em minus 0.4em\relax IEEE, 2023, pp. 1--14.

\bibitem{chainalysis2023}
\BIBentryALTinterwordspacing
{Chainalysis}, ``The 2023 crypto crime report,'' 2023, accessed: Nov. 19, 2025. [Online]. Available: \url{https://www.chainalysis.com/blog/2023-crypto-crime-report-introduction}
\BIBentrySTDinterwordspacing

\bibitem{worldbank2024remittance}
\BIBentryALTinterwordspacing
{World Bank}, ``Remittance prices worldwide quarterly report (issue 49),'' The World Bank Group, Tech. Rep., March 2024. [Online]. Available: \url{https://documents.worldbank.org/en/publication/documents-reports/documentdetail/099053025155726868}
\BIBentrySTDinterwordspacing

\bibitem{ffnews2024}
\BIBentryALTinterwordspacing
F.~News, ``Dlocal partners with moneygram to expand leading international payment services in high growth markets,'' 2024, accessed: 24 Dec, 2025. [Online]. Available: \url{https://ffnews.com/newsarticle/paytech/dlocal-partners-with-moneygram-to-expand-leading-international-payment-services-in-high-growth-markets}
\BIBentrySTDinterwordspacing

\bibitem{jhanji2025stablecoins}
\BIBentryALTinterwordspacing
K.~Jhanji, K.~Burchardi, Y.~H. Zhang, C.~Bravo, T.~Hung, B.~Kronfellner, and H.~Samad, ``Stablecoins: Five killer tests to gauge their potential,'' Boston Consulting Group, White Paper, May 2025. [Online]. Available: \url{https://media-publications.bcg.com/Stablecoins-five-killer-tests-to-gauge-their-potential.pdf}
\BIBentrySTDinterwordspacing

\bibitem{buterin2021erc4337}
\BIBentryALTinterwordspacing
V.~Buterin, Y.~Weiss, D.~Tirosh \emph{et~al.}, ``Erc-4337: Account abstraction using alt mempool,'' 2021. [Online]. Available: \url{https://eips.ethereum.org/EIPS/eip-4337}
\BIBentrySTDinterwordspacing

\bibitem{worldpay2022usdc}
\BIBentryALTinterwordspacing
{Worldpay}, ``Usdc role in payment landscape grows,'' 2022, accessed: Nov. 27, 2025. [Online]. Available: \url{https://www.worldpay.com/en/insights/articles/usdc-stablecoin-circle}
\BIBentrySTDinterwordspacing

\bibitem{nuvei2024crypto}
\BIBentryALTinterwordspacing
{Nuvei}, ``Nuvei launches comprehensive blockchain payment solution for cross-border b2b payments,'' December 2024, accessed: Nov. 27, 2025. [Online]. Available: \url{https://www.nuvei.com/posts/nuvei-launches-comprehensive-blockchain-payment-solution}
\BIBentrySTDinterwordspacing

\bibitem{paypal2023pyusd}
\BIBentryALTinterwordspacing
{PayPal}, ``Paypal launches u.s. dollar stablecoin,'' August 2023, accessed: Nov. 27, 2025. [Online]. Available: \url{https://newsroom.paypal-corp.com/2023-08-07-PayPal-Launches-U-S-Dollar-Stablecoin}
\BIBentrySTDinterwordspacing

\bibitem{visa2023stablecoin}
\BIBentryALTinterwordspacing
{Visa}, ``Visa expands stablecoin settlement capabilities to merchant acquirers,'' September 2023, accessed: Nov. 27, 2025. [Online]. Available: \url{https://usa.visa.com/about-visa/newsroom/press-releases.releaseId.19881.html}
\BIBentrySTDinterwordspacing

\bibitem{chainlink2021ccip}
\BIBentryALTinterwordspacing
L.~Breidenbach \emph{et~al.}, ``Chainlink 2.0: Next steps in the evolution of decentralized oracle networks,'' Chainlink Labs, Tech. Rep., 2021. [Online]. Available: \url{https://research.chain.link/whitepaper-v2.pdf}
\BIBentrySTDinterwordspacing

\bibitem{belchior2021survey}
R.~Belchior, A.~Vasconcelos, S.~Guerreiro, and M.~Correia, ``A survey on blockchain interoperability: Past, present, and future trends,'' \emph{ACM Computing Surveys}, vol.~54, no.~8, pp. 1--41, 2021.

\bibitem{bis2023blueprint}
\BIBentryALTinterwordspacing
{Bank for International Settlements}, ``Annual economic report 2023. chapter iii: Blueprint for the future monetary system: improving the old, enabling the new,'' Tech. Rep., 2023. [Online]. Available: \url{https://www.bis.org/publ/arpdf/ar2023e3.htm}
\BIBentrySTDinterwordspacing

\bibitem{CPMI-IOSCO-2022-Stablecoin}
\BIBentryALTinterwordspacing
------, ``Application of the pfmi to stablecoin arrangements: Final report and accompanying report,'' accessed: Oct. 10, 2025. [Online]. Available: \url{https://www.bis.org/cpmi/publ/d207.htm}
\BIBentrySTDinterwordspacing

\bibitem{auer2022central}
R.~Auer, J.~Frost, L.~Gambacorta, C.~Monnet, T.~Rice, and H.~S. Shin, ``Central bank digital currencies: motives, economic implications, and the research frontier,'' \emph{Annual review of economics}, vol.~14, no.~1, pp. 697--721, 2022.

\bibitem{prodan2024rise}
S.~Prodan, P.~Konh{\"a}usner, D.-C. Dabija, G.~Lazaroiu, and L.~Marincean, ``The rise in popularity of central bank digital currencies. a systematic review,'' \emph{Heliyon}, vol.~10, no.~9, 2024.

\bibitem{bis2024tokenisation}
\BIBentryALTinterwordspacing
{Bank for International Settlements}, ``Tokenisation in the context of money and other assets,'' 2024. [Online]. Available: \url{https://www.bis.org/cpmi/publ/d225.pdf}
\BIBentrySTDinterwordspacing

\bibitem{al2023anti}
T.~N. Al-Tawil, ``Anti-money laundering regulation of cryptocurrency: Uae and global approaches,'' \emph{Journal of Money Laundering Control}, vol.~26, no.~6, pp. 1150--1164, 2023.

\end{thebibliography}

\end{document}